\documentclass[11pt]{amsart}
\usepackage{amsbsy,amssymb,amsmath,amsthm,amscd,amsfonts,latexsym,amstext,delarray,
amsmath,graphicx} 
\usepackage[margin=1in]{geometry}
\usepackage{color}
\input xypic

\usepackage[all]{xy}
\usepackage{setspace}

\newtheorem{thm}{Theorem}[section]
\newtheorem{prop}[thm]{Proposition}
\newtheorem{cor}[thm]{Corollary}
\newtheorem{lem}[thm]{Lemma}

\newtheorem{rem}[thm]{Remark}
\newtheorem{ex}[thm]{Example}

\numberwithin{equation}{section}

\def\F{{\mathbb F}}
\def\Q{{\mathbb Q}}
\def\Z{{\mathbb Z}}

\def\R{{\mathbb R}}
\def\C{{\mathbb C}}
\def\H{{\mathbb H}}

\def\P{{\mathbb P}}

\def\PSL{{\rm PSL}}
\def\SL{{\rm SL}}
\def\Sym{{\rm Sym}}

\def\cA{{\mathcal A}}

\def\cD{{\mathcal D}}
\def\cE{{\mathcal E}}
\def\cF{{\mathcal F}}
\def\cG{{\mathcal G}}
\def\cH{{\mathcal H}}

\def\cK{{\mathcal K}}
\def\cL{{\mathcal L}}
\def\cM{{\mathcal M}}

\def\cO{{\mathcal O}}
\def\cP{{\mathcal P}}

\def\cR{{\mathcal R}}
\def\cS{{\mathcal S}}

\def\cV{{\mathcal V}}
\def\cW{{\mathcal W}}

\def\bK{{\mathbb K}}

\def\bS{{\mathbb S}}
\def\bT{{\mathbb T}}

\def\tr{{\rm tr}}

\def\Ind{{\rm Ind}}
\def\Conf{{\rm Conf}}

\DeclareMathOperator*{\Hom}{Hom}

\def\cancel#1#2{\ooalign{$\hfil#1\mkern1mu/\hfil$\crcr$#1#2$}}

\def\dirac{\mathpalette\cancel\partial}

\title[Twisted Index, Orbifold Symmetric Products, and FQHE]{Twisted index theory on 
orbifold symmetric products and the fractional quantum Hall effect}
\author{Matilde Marcolli and Kyle Seipp}
\address{Mathematics Department, Caltech, 1200 E. California Blvd. Pasadena, CA 91125, USA}
\email{matilde@caltech.edu}
\email{kseipp@caltech.edu}
\date{}

\begin{document}
\maketitle

\begin{abstract}
We extend the noncommutative geometry model of the fractional quantum Hall effect,
previously developed by Mathai and the first author, to orbifold symmetric products. 
It retains the same properties of quantization of the Hall conductance
at integer multiples of the fractional Satake orbifold Euler characteristics. We show that it 
also allows for interesting composite fermions and anyon representations, and possibly 
for Laughlin type wave functions.
\end{abstract}


\section{Introduction}

A satisfactory model of the integer quantum Hall effect within the framework of noncommutative geometry was
developed in \cite{Bell}, \cite{Bell2}. In a $2$-dimensional periodic lattice, 
the presence of an external magnetic field turns
the classical Brillouin zone into a non-commutative torus, replacing the ordinary translational symmetries of
the Hamiltonian by magnetic translations. These are symmetries of the magnetic Laplacian, 
and they only commute up to a phase factor, hence the appearance of
the noncommutative torus as the new algebra of observables. 
The integer quantization of the Hall conductance can then be interpreted in terms of an index theorem on the 
noncommutative torus.

\smallskip

In \cite{MaMa1}, \cite{MaMa2}, \cite{MaMa3} a single particle model was developed for a charged particle moving in a magnetic 
field within a curved geometry described by a good 2-dimensional orbifold, with the curved
geometry simulating an averaged effect of the interaction with other particles.
This model exhibits quantization of the Hall conductance
at fractional values given by integer multiples of the Satake orbifold Euler characteristic. The results of 
\cite{MaMa1}, \cite{MaMa2}, \cite{MaMa3} are based on a
generalization to the (fractional) orbifold case of a previous treatment, in \cite{CHMM}, of the integer quantum 
Hall effect in the hyperbolic geometry of a smooth Riemann surfaces of genus $g\geq 2$.

\smallskip

The main drawback of this noncommutative-geometric approach to the fractional quantum Hall
effect lies in the fact that it is still based on a single particle model. While the integer quantum
Hall effect is well described by an independent electron approximation, which reduces it to
a single particle model, the fractional quantum Hall effect is intrinsically a many particle phenomenon:
while the single particle model used in \cite{MaMa1}, \cite{MaMa2}, \cite{MaMa3} produces a
fractional quantization of the Hall conductance as a Kawasaki orbifold index-theorem on the
relevant noncommutative space, it does not account for Laughlin type wave functions, nor for
composite fermion (or anyon) representations.  

\smallskip

In this paper we propose a way to extend the noncommutative geometry model of 
\cite{MaMa1}, \cite{MaMa2}, \cite{MaMa3}, so that it makes contact with field theories
on orbifolds, of the kind considered in relation to String Theory, see e.g.~\cite{Adem},
\cite{DHVW}, \cite{MooreRead}, \cite{VaWi}.
We consider systems of $n$ indistinguishable particles moving in the same type of homogeneous 
negatively curved geometry, under the effect of an external magnetic field, so that a classical
configuration of the system is described by a point in the $n$-fold symmetric products 
of a good 2-dimensional orbifold. 
As in the case of field theories on orbifolds, the relevant Fock space is given by the
sum of the orbifold K-theories (or the delocalized equivariant cohomology) of the twisted 
group $C^*$-algebras of the wreath products $\Gamma_n=\Gamma^n\rtimes S_n$, with
$\Gamma$ the orbifold fundamental group of the good 2-dimensional orbifold. The quantization
of the Hall conductance is still obtained via a twisted higher index theorem as in \cite{MaMa2}
and is expressible in terms of the Satake orbifold Euler characteristics of the orbifold symmetric
products. At the same time, the model now allows for interesting composite fermion and
anyon representations, whose classification depends on Seifert invariants of orbifold line
bundles. We also formulate some hypothesis, still speculative at this stage, on how to
obtain Laughlin type wave functions from the geometry of the model.

\smallskip

The paper is structured as follows: in the rest of this introductory section we discuss the
geometry of $2$-dimensional good orbifolds $\Sigma$ and their symmetric 
products $\Sym^n(\Sigma)$. 
In Section \ref{pi1orbSec}, we introduce the 
relevant groups that we will be considering, related to various kinds of orbifold covers.
In particular, we extend to orbifold fundamental groups a result for smooth Riemann 
surfaces, which identifies the (orbifold) fundamental group of the 
symmetric products $\Sym^n(\Sigma)$ with
the abelianization of the (orbifold) fundamental group of $\Sigma$. In Section \ref{KtheorySec},
we focus on K-theoretic aspects. We compute the orbifold $K$-theory of $\Sym^n(\Sigma)$,
in terms of classifying spaces for proper actions and we relate it to the $K$-theory of
the group $C^*$-algebras $C^*_r(\Gamma_n)$ via the Baum--Connes conjecture, which
we show is satisfied by the wreath products $\Gamma_n$. We also discuss a possible 
notion of orbifold-Jacobian and its K-theoretic properties. In Section \ref{magfieldSec},
we show that the magnetic field determines a compatible family of $U(1)$-multipliers $\sigma_n$
on $\Gamma_n$. We obtain in this way
twisted group $C^*$-algebras $C^*_r(\Gamma_n,\sigma_n)$ generalizing
the algebra $C^*_r(\Gamma,\sigma)$ considered in \cite{MaMa1}, \cite{MaMa2}, \cite{MaMa3}.
Using the Lyndon--Hochschild--Serre spectral sequence for the group cohomology of the 
wreath products, we show that the multipliers $\sigma_n$ define cocycles with trivial 
Dixmiar--Douady class. We then show that the $K$-theory of the twisted group 
$C^*$-algebra $C^*_r(\Gamma_n,\sigma_n)$ agrees with the $K$-theory of the
untwisted $C^*_r(\Gamma_n)$. To this purpose, we prove the K-amenability of
the wreath products $\SL(2,\R)^n\rtimes S_n$, by adapting the argument of \cite{FoxHask}
for the K-amenability of $\SL(2,\R)$. Section \ref{EulerSec} contains expository material,
where we recall and compare the different notions of orbifold Euler characteristic used
in the Kawasaki index theorem and in string theory on orbifolds. The former, which we
refer to as the Satake orbifold Euler characteristic is in general a rational number,
while the latter, which we call the string-theoretic orbifold Euler characteristic is an integer. 
We recall how the latter relates to sectors and to inertia orbifolds, and how it generalizes
to orbifold Chern classes. We also recall the Segal construction of the Fock space for
orbifold symmetric products, based on equivariant K-theory, as in \cite{Segal}, \cite{Wang}.
In Section \ref{IndexSec} we compute the twisted higher index theorem of \cite{MaMa2}
for the orbifold symmetric products $\Sym^n(\Sigma)$. Using the
same relation between Hall conductance cocycle and area cocycle as in
\cite{CHMM}, \cite{MaMa2}, we show that the Hall conductance is quantized
at fractional values equal to integer multiples of the Satake orbifold Euler 
characteristic $\chi^{orb}(\Sym^n(\Sigma))$. In Section \ref{AnyonsSec} we 
classify composite fermions and anyons on the symmetric products $\Sym^n(\Sigma)$.
We introduce a notion of orbifold braid group, which is the orbifold fundamental group
of the configuration spaces ${\rm Conf}(\Sigma,n)$, where the orbifold singularities
are coming from the orbifold cone points of $\Sigma$. We show that, similarly to
what happens in the case of Riemann surfaces and ordinary braid groups, if the orbifold
$\Sigma$ has genus $g>0$, then the scalar unitary representations can only be fermions
or bosons, with no non-trivial anyons. When the genus is $g=0$, there are anyon representations
and we show that they are classified by the Seifert invariants of an orbifold line bundle
with integer orbifold Euler number. We also show that, for arbitrary genus, there are 
anyon representations of higher dimensions $N$, which again depend on Seifert invariants,
for an orbifold line bundle whose orbifold Euler number is in $\Z + (g+n-1)/N$, and with
fractional statistics $\pi i/N$. Finally Section \ref{LaughlinSec} contains some more
speculative considerations on how to find Laughlin type wave functions in this geometric
setting, in terms of the Mathai--Quillen formalism for Euler classes of vector bundles,
and local systems determined by a given $N$-dimensional anyon representation. The
relation between $N$ and $\# G$ imposed by the classification of anyon representations
implies that one finds powers of the Vandermonde determinant with exponents 
equal to the denominators that appear in the quantization of the Hall conductance,
as expected in Laughlin wave functions. We also suggest the possibility that 
Laughlin type functions may
appear in computations via Selberg integrals of the orbifold Euler characteristic
of a moduli space of good $2$-dimensional orbifolds, analogous to the known
calculations for moduli spaces of curves with marked points.

\bigskip

In the rest of this introductory section we review some known material that we need in the following.

\subsection{Hyperbolic $2$-dimensional good orbifolds}

Let $\H$ denote the $2$-dimensional hyperbolic plane. We will use either the 
upper half plane model $\H \cong \{ z=x+iy \in \C \,|\, \Im(z)>0 \}$ with the metric $ds^2 = (dx^2 + dy^2)/y^2$
or the equivalent Poincar\'e disc model $\H \cong \{ z\in \C \,|\, |z|<1 \}$, with the metric
$ds^2 =4 (dx^2 + dy^2)/(1-|z|^2)^2$.

\smallskip

On $\H$ we consider the isometric action of a discrete cocompact subgroup $\Gamma \subset \PSL(2,\R)$,
given by a Fuchsian group of signature $(g,\underline{\nu})$ with $\underline{\nu}=(\nu_1,\ldots,\nu_m)$. These groups have an
explicit presentation with generators $a_i,b_i$ with $i=1,\ldots, g$ and $c_j$ with $j=1,\ldots, m$, of the form
\begin{equation}\label{Fuchsian}
\Gamma=\Gamma(g,\underline{\nu})=\langle a_i, b_i, c_j\,|\, \prod_{i=1}^g [a_i,b_i] c_1 \cdots c_m =1, \ c_j^{\nu_j}=1 \rangle.
\end{equation}
The quotient $\Sigma=\Sigma(g,\underline{\nu})=\H/\Gamma$ is a $2$-dimensional hyperbolic good orbifold. 
It is a Riemann surface of genus $g$ with $m$ cone points $\{ x_1, \ldots, x_m \}$, where the point 
$x_j$ has stabilizer of order $\nu_j$.
Any such orbifold has a finite branched covering by a smooth Riemann surface $\Sigma_{g'}$, with
$\Sigma(g,\underline{\nu})=\Sigma_{g'}/G$ for a finite group $G$. The genus $g'$ is related to $g$ by 
the Riemann--Hurwitz formula for branched coverings:
\begin{equation}\label{genusrel}
 g' = 1 + \frac{\# G}{2} (2(g-1) + (m-\sum_j \nu_j^{-1})). 
\end{equation} 
The Riemann surface $\Sigma_{g'}=\H/\Gamma'$ has a hyperbolic uniformization by $\Gamma' \subset \PSL(2,\R)$,
which is related to $\Gamma(g,\underline{\nu})$ by an exact sequence
\begin{equation}\label{GammaGseq}
 1 \longrightarrow \Gamma_{g'} \longrightarrow \Gamma(g,\underline{\nu}) \longrightarrow G \longrightarrow 1. 
\end{equation} 
The Fuchsian group $\Gamma= \Gamma(g,\underline{\nu})$ is the {\em orbifold fundamental group} of the $2$-dimensional orbifold $\Sigma=\Sigma(g,\underline{\nu})$, see \cite{Scott}. 

\smallskip
\subsection{Symmetric products}

Let $\Sym^n(X)=X^n/S_n$, with $S_n$ the group of permutations of a set of $n$ elements. It is
well known that, for a smooth compact Riemann surface $\Sigma$ of genus $g$, the symmetric products
$\Sym^n(\Sigma)$ are smooth and are related to the Jacobian of $\Sigma$ in the following way.
Let $\Omega^1(\Sigma)$ be the space of holomorphic $1$-forms with a basis $\{ \omega_1, \ldots, \omega_g \}$.
The group $\cP_\Sigma$ of periods of $\Sigma$ is the subgroup $\cP_\Sigma \subset \C^g$ given by
the $v=(v_1,\ldots,v_g)\in \C^g$ obtained as integrals $v_i =\int_\gamma \omega_i$ for some
$\gamma \in \pi_1(\Sigma)$. For a base point $x_0 \in \Sigma$, and a path $\gamma'$ from $x_0$ to $x\in \Sigma$,
the integral $\int_{\gamma'} \omega_i$ then defines the Abel--Jacobi map 
\begin{equation}\label{AJmap}
\cA: \Sigma \to J(\Sigma)=\C^g/\cP_\Sigma, \ \ \  \cA: x \mapsto \left(\int_{x_0}^x \omega_1, \ldots, \int_{x_0}^x \omega_g \right),
\end{equation}
where the Jacobian $J(\Sigma)$ is a torus $T^{2g}$, which can also be identified with $H^1(\Sigma,\R)/H^1(\Sigma,\Z)$.
The Abel--Jacobi map extends to a map 
\begin{equation}\label{AJsym}
\cA: \Sym^n(\Sigma) \to J(\Sigma), \ \ \ \cA: [x_1,\ldots, x_n] \mapsto \cA(x_1)+\cdots+\cA(x_n) .
\end{equation}
When $n> 2g-2$, the symmetric products fiber over the Jacobian with fibers that are projective spaces of dimension $n-g$,
\begin{equation}\label{JacSym}
\P^{n-g}(\C) \hookrightarrow \Sym^n(\Sigma) \twoheadrightarrow J(\Sigma) .
\end{equation}

\smallskip

Moreover, it was shown in Remark 5.8 of \cite{SGA} (see also \cite{KaTa} for a generalization) that
the fundamental group of the symmetric products of a smooth compact Riemann surface satisfies, for all $n\geq 2$,
\begin{equation}\label{fundgrsymRS}
\pi_1 (\Sym^n(\Sigma)) = \pi_1(\Sigma)^{ab} = H^1(\Sigma,\Z) = \pi_1(J(\Sigma)).
\end{equation}

\smallskip

In the following section we will consider the symmetric products $\Sym^n(\Sigma)$ of a $2$-dimensional
hyperbolic orbifold $\Sigma=\Sigma(g,\underline{\nu})$ and we compute the orbifold fundamental group.

\section{Symmetric products, orbifold fundamental group, and orbifold coverings}\label{pi1orbSec}

In this section we discuss various orbifold coverings of the symmetric products $\Sym^n(\Sigma)$
and their associated groups of symmetries.

\begin{lem}\label{orbXn}
Let $\Sigma$ be a good $2$-dimensional orbifold, with singular locus $\Sigma_{\rm sing}$ given by
a finite set of cone points, and with orbifold fundamental group $\pi_1^{orb}(\Sigma)$. Then
\begin{equation}\label{pi1orbXn}
\pi_1^{orb}(\Sigma^n) \cong \pi_1^{orb}(\Sigma)^n. 
\end{equation}
\end{lem}

\proof Let $\cO$ be a good orbifold, with singular locus of (real) codimension two, 
${\rm codim} \cO_{\rm sing} = 2$.
Then the orbifold fundamental group $\pi_1^{orb}(\cO)$ of an orbifold $\cO$ 
can be described (\cite{Thur}, \S 13) as the quotient 
\begin{equation}\label{pi1orbQuot}
\pi_1^{orb}(\cO) = \pi_1(\cO_{\rm reg})/ H,
\end{equation}
of the fundamental group $\pi_1(\cO_{\rm reg})$
of the regular part $\cO_{\rm reg}=\cO\smallsetminus \cO_{\rm sing}$ of the orbifold
(the complement of the singular locus) by the normal subgroup $H$ generated by 
the classes $\gamma_j^{\nu_j}$ in $\pi_1(\cO_{\rm reg})$, where $\gamma_j$ are
loops around a component $\cO_j$ of $\cO_{\rm sing}$ with $\nu_j$ the order of the
stabilizer $G_j$ of $\cO_j$. In particular, for a $2$-dimensional good orbifold $\Sigma$
we have $\Sigma_{\rm sing}=\{ x_j \}_{j=1,\ldots, m}$ the cone points with stabilizers 
$\Z/\nu_j\Z$. In the product $\Sigma^n$ we have $(\Sigma^n)_{\rm sing}=\cup_{k=1}^n 
\Sigma_{{\rm sing},(k)}$, where  $\Sigma_{{\rm sing},(k)}$ means a copy of 
$\Sigma_{\rm sing}$ in the $k$-th factor and the full space $\Sigma$ in all the other factors 
$\Sigma_{{\rm sing},(k)}= \Sigma\times \cdots \times \Sigma_{{\rm sing}} \times \cdots \times \Sigma$.
Thus, the set of regular points $(\Sigma^n)_{\rm reg} = \Sigma^n \smallsetminus
\Sigma^n_{\rm sing}$ is given by $(\Sigma^n)_{\rm reg} =(\Sigma_{\rm reg})^n$, and 
we can unambiguously use the notation $\Sigma^n_{\rm reg}$ for this locus.
We have $\pi_1(\Sigma^n_{\rm reg})=\pi_1(\Sigma_{\rm reg})^n$. The normal subgroup
$H$ of $\pi_1(\Sigma^n_{\rm reg})$ is generated by loops $\gamma_{j,k}$ in $\Sigma^n$ 
that circle around the components $\Sigma \times \cdots \times \{ x_j \} \times \cdots \times \Sigma$
of $\Sigma_{{\rm sing},k}$. It suffices then to observe that for $k\neq k'$ and for all $j,j'$, the
elements $\gamma_{j,k}$ and $\gamma_{j',k'}$ commute in $\pi_1(\Sigma^n_{\rm reg})$,
so that the group $H$ is a direct product $H=\prod_{k=1}^n H_k$ with each $H_k$
isomorphic to the normal subgroup of the $k$-th factor $\pi_1(\Sigma_{\rm reg})$, with
$\pi_1(\Sigma_{\rm reg})/H_k =\pi_1^{orb}(\Sigma)$. Thus, we obtain 
$$ \pi_1^{orb}(\Sigma^n)= \pi_1(\Sigma^n_{\rm reg})/H = \pi_1(\Sigma_{\rm reg})^n/ \prod_k H_k =
\pi_1^{orb}(\Sigma)^n. $$
\endproof

\bigskip

Let $\Gamma$ be a discrete group. Let $\Gamma^n\rtimes S_n$ be the semidirect product
with multiplication $$(g_1,\ldots,g_n,\sigma)(h_1,\ldots,h_n,\tau)= (g_1 h_{\sigma(1)}, \ldots, g_n h_{\sigma(n)}, \sigma\tau),$$
and let $\langle S_n \rangle \subset \Gamma^n\rtimes S_n$ be the normal subgroup generated by the elements of $S_n$.

The following argument is implicit in Remark 5.8 of \cite{SGA}. We reformulate it
here in purely topological terms.

\begin{lem}\label{pi1orbSn}
Let $\Sigma$ be a good $2$-dimensional orbifold, with singular locus $\Sigma_{\rm sing}$ given
by a finite set of cone points. Then the orbifold fundamental groups of the symmetric 
products $\Sym^n(\Sigma)$ satisfy
\begin{equation}\label{pi1orbSymn}
\pi_1^{orb}(\Sym^n(\Sigma)) \cong \pi_1^{orb}(\Sigma)^n \rtimes S_n / \langle S_n \rangle.
\end{equation}
\end{lem}

\proof As in \eqref{pi1orbQuot} in the previous Lemma, we have $\pi_1^{orb}(\Sigma) = 
\pi_1(\Sigma_{\rm reg})/ H$.
The ordinary fundamental group $\pi_1(\Sigma_{\rm reg})$ classifies (ordinary) 
covering spaces of $\Sigma_{\rm reg}$,
in the sense that, to each normal subgroup $N$ of $\pi_1(\Sigma_{\rm reg})$ there corresponds a regular covering
space $\Sigma_N$ of $\Sigma_{\rm reg}$. Such a covering space is a principal 
$\pi_1(\Sigma_{\rm reg})/N$-fibration over
$\Sigma_{\rm reg}$. In particular, the quotient 
$\pi_1^{orb}(\Sigma) = \pi_1(\Sigma_{\rm reg})/ H$ similarly classifies all such
coverings of $\Sigma_{\rm reg}$ that extend to a branched covering of $\Sigma$ with a 
trivial action of the stabilizers
of the singular points on the corresponding fibers. Such coverings correspond to normal subgroups $N$ of
$\pi_1(\Sigma_{\rm reg})$ that contain the normal subgroup $H$ generated by loops $\gamma_j$ around the 
components of $\Sigma_{\rm sing}$ with the appropriate multiplicities $\nu_j$ as above. Next observe that a
regular covering $\Sigma_N$ of $\Sigma_{\rm reg}$ can also be described as an $S_n$-equivariant covering
$\tilde\Sigma_N$ of $\Sigma_{\rm reg}^n \smallsetminus \Delta$, the complement of the diagonals in 
$(\Sigma_{\rm reg})^n=(\Sigma^n)_{\rm reg}$. This means that there is an action of $S_n$ on $\tilde\Sigma_N$,
which is compatible with the action on $\Sigma_{\rm reg}^n \smallsetminus \Delta$, so that the projection
map of the covering is $S_n$-equivariant. These are classified by the crossed product 
$\pi_1(\Sigma_{\rm reg}^n \smallsetminus \Delta) \rtimes S_n$ of the action of $S_n$ on the 
fundamental group $\pi_1(\Sigma_{\rm reg}^n \smallsetminus \Delta)$. 
Among all such coverings, we consider those that extend 
to an $S_n$-equivariant branched covering of $\Sigma^n$, with a trivial action of the stabilizers of the components
of $\Sigma_{\rm sing}$ and of the diagonals. These are then classified by a quotient of 
$\pi_1(\Sigma_{\rm reg}^n \smallsetminus \Delta) \rtimes S_n$ where we mod out by the normal
subgroup generated by the loops around the components of $\Sigma_{\rm sing}$ and 
the elements of $S_n$, that is, by the group $\pi_1^{orb}(\Sigma^n)\rtimes S_n /\langle S_n \rangle$.
Finally, observe that the data of an $S_n$-equivariant branched covering of $\Sigma^n$ as above
uniquely determine a branched covering over the symmetric product $\Sym^n(\Sigma)$ and vice
versa,
so that we can identify $\pi_1^{orb}(\Sigma^n)\rtimes S_n /\langle S_n \rangle = \pi_1^{orb}(\Sym^n(\Sigma))$.
\endproof

\bigskip

The following observation is also implicit in Remark 5.8 of \cite{SGA}. We spell it out for convenience.

\begin{lem}\label{Gab}
Let $\Gamma^n\rtimes S_n$ be as above, with $\langle S_n \rangle \subset \Gamma^n\rtimes S_n$ 
the normal subgroup generated by the elements of $S_n$.
There is a group isomorphism
\begin{equation}\label{SnGab}
\Gamma^n\rtimes S_n / \langle S_n \rangle \cong \Gamma^{ab},
\end{equation}
where $\Gamma^{ab}=\Gamma/[\Gamma,\Gamma]$ is the abelianization.
\end{lem}

\proof Let $g_{(i)}$ denote the element $g_{(i)}=(1,\ldots, 1, g, 1, \ldots, 1)$
of $\Gamma^n$ that has the $i$-th coordinate equal to $g\in \Gamma$
and all the other coordinates equal to the identity element $1$. 
Since $\Gamma^n$ is a direct product of copies of $\Gamma$, the different factors $\Gamma$ 
commute with each other, hence for all $g,h\in \Gamma$,
we have $g_{(i)}h_{(j)}=h_{(j)} g_{(i)}$ whenever $i\neq j$, with the product equal to the  
element of $\Gamma^n$ with $g$ in the $i$-th place, $h$ in the $j$-th place and $1$ everywhere else.
Moreover, observe that, for any $g\in \Gamma$ and for $i\neq j$, the element $(g_{(i)},1)$ in $\Gamma^n\rtimes S_n$
is equal to $(1,\sigma_{ij})^{-1} (g_{(j)},1) (1,\sigma_{ij})$, where $\sigma_{ij}\in S_n$ is the permutation that
exchanges $i$ and $j$ and fixes all other elements of $\{ 1, \ldots, n\}$. For any $\gamma\in \Gamma^n\rtimes S_n$,
and for all $\sigma \in S_n$, we have $(1,\sigma)^{-1} \gamma (1,\sigma) \gamma^{-1} \in \langle S_n \rangle$,
hence $\gamma$ and $(1,\sigma)^{-1} \gamma (1,\sigma)$ define the same class in the quotient
$\Gamma^n\rtimes S_n/\langle S_n \rangle$. In particular, $g_{(i)}$ and $g_{(j)}$ define the same element
in the quotient, for all $g\in \Gamma$ and for all $i\neq j$. Thus, we obtain that in the quotient the $n$-copies of
$\Gamma$ in the product $\Gamma^n$ are identified and 
commutators are killed, hence the quotient gets
identified with $\Gamma^{ab}$.
\endproof

\bigskip

Combining the results of Lemmata \ref{orbXn}, \ref{pi1orbSn}, and \ref{Gab}, we obtain the analog of \eqref{fundgrsymRS}
for orbifold fundamental groups.

\begin{prop}\label{mainpi1orb}
For $n\geq 2$, the symmetric products $\Sym^n(\Sigma)$ of a good $2$-dimensional 
orbifold $\Sigma$ have orbifold fundamental group given by
\begin{equation}\label{pi1orbab}
\pi_1^{orb}(\Sym^n(\Sigma)) \cong \pi_1^{orb}(\Sigma)^{ab}.
\end{equation}
\end{prop}

\bigskip

In the case of a $2$-dimensional hyperbolic orbifold $\Sigma=\Sigma(g,\underline{\nu})$ one obtains the following.

\begin{cor}\label{2dorbSn}
For $\Sigma=\Sigma(g,\underline{\nu})$ and $n\geq 2$, we have $\pi_1^{orb}(\Sym^n(\Sigma(g,\underline{\nu})))=\Z^{2g}\oplus \Z_\nu$,
where $\Z_\nu=\oplus_{j=1}^m \Z/\nu_j\Z$.
\end{cor}

\proof The abelianization of a group $\Gamma=\Gamma(g,\underline{\nu})$ of the form \eqref{Fuchsian} is given by $\Z^{2g}\oplus_j \Z/\nu_j\Z$.
\endproof

\medskip

\begin{rem}\label{H1orb}{\rm The abelianization of the fundamental group is the first homology group,
$\pi_1(\Sigma)^{ab}=H^1(\Sigma,\Z)$. In the case of the orbifold fundamental group, there is a similar
homological interpretation of its abelianization, in terms of the $t$-singular homology defined in \cite{TaYo},
$\pi_1^{orb}(\Sigma)^{ab}=tH^1(\Sigma,\Z)$, where the $t$-singular homology $tH^*$ is constructed using
singular simplexes that intersect transversely the singular locus of $\Sigma$, see \cite{TaYo} for details.}
\end{rem}

\subsection{Geometry of some orbifold covering spaces}

We consider some covering spaces of the orbifold symmetric products that will be useful in the
rest of the paper.

\begin{prop}\label{orbcoverings}
For $n\geq 2$, let $\Sym^n(\Sigma)$ be the symmetric product of the $2$-dimensional orbifold 
$\Sigma=\Sigma(g,\underline{\nu})$. Let $\langle S_n \rangle$ be the normalizer of $S_n$ in $\Gamma^n\rtimes S_n$
and let $\bS_n =\langle S_n \rangle/S_n$. Let $\bS^n(\H):=\H^n/\langle S_n \rangle$. Let $\Sigma_{g'}=\H/\Gamma_{g'}$
with $\Gamma_{g'}$ as in \eqref{GammaGseq} with finite quotient $G=\Gamma/\Gamma_{g'}$. Let $G_n:=G^n \rtimes S_n$.
\begin{enumerate}
\item $\Sym^n(\Sigma)$ is orbifold covered by $\H^n$, 
with $\Sym^n(\Sigma) = \H^n / \Gamma^n \rtimes S_n$.
\item $\Sym^n(\Sigma)$ is orbifold covered by $\bS^n(\H)$, with 
$\Sym^n(\Sigma)= \bS^n(\H)/\pi_1^{orb}(\Sym^n(\Sigma))$.
\item $\bS^n(\H)$ is orbifold covered by 
$D^{2n}=\Sym^n(\H)$, with $\bS^n(\H)=D^{2n}/\bS_n$.
\item $\Sym^n(\Sigma)$ is orbifold covered by the smooth manifold
$\Sigma_{g'}^n$, with $\Sym^n(\Sigma)=\Sigma_{g'}^n/G_n$. 
\end{enumerate}
\end{prop}

\proof
The orbifold $\Sigma=\Sigma(g,\underline{\nu})$ has a finite branched cover by a smooth surface
$\Sigma_{g'}$, of genus \eqref{genusrel}, so that $\Sigma=\Sigma_{g'}/G$, with 
the finite group $G$ as in \eqref{GammaGseq}. Moreover, $\Sigma$ also has universal
orbifold cover $\H$, with $\Sigma=\H/\Gamma$, for $\Gamma=\Gamma(g,\underline{\nu})$.
The isometric action of $\Gamma$ on $\H$ induces an isometric action of $\Gamma^n\rtimes S_n$ on
the $n$-fold product $\H^n$, with quotient $\H^n/\Gamma^n\rtimes S_n =\Sym^n(\Sigma)$. 
Consider the normal subgroup $\langle S_n \rangle \subset \Gamma^n\rtimes S_n$. We can equivalently
describe the quotient above as $(\H^n/\langle S_n \rangle)/(\Gamma^n\rtimes S_n/\langle S_n \rangle) 
=\Sym^n(\Sigma)$. The group $S_n$ is normal inside $\langle S_n \rangle$ with quotient $\bS_n$,
and we can further write the quotient $\H^n/\langle S_n \rangle=(\H^n/S_n)/\bS_n=\Sym^n(\H)/\bS_n$.
We identify the hyperbolic plane $\H$ with its Poincar\'e disc model $\H=D^2$ (the open
unit disc in $\R^2$ with the hyperbolic metric). By Lemma 5 of \cite{KaSa}, there is a homeomorphism of pairs
$$ (D^{2n}, \partial D^{2n}=S^{2n-1}) \cong (\Sym^n(D^2), \Sym^n(\overline{D^2})\smallsetminus \Sym^n(D^2)), $$
where $D^{2n}$ is an open $2n$-dimensional disc.
Thus, we can identify $\Sym^n(\H)\cong D^{2n}$, with the metric induced by the hyperbolic metric on $\H$,
so that $\Sym^n(\H)/\bS_n=D^{2n}/\bS_n$.
Finally, consider the sequence of groups \eqref{GammaGseq}. The normal embedding $\Gamma_{g'} \hookrightarrow \Gamma$
determines a normal embedding $\Gamma_{g'}^n \hookrightarrow \Gamma^n\rtimes S_n$. The quotient group can be
identified with $G^n\rtimes S_n$, where $G=\Gamma/\Gamma_{g'}$. We then rewrite the quotient 
$\H^n/\Gamma^n\rtimes S_n =\Sym^n(\Sigma)$ as $(\H^n/\Gamma_{g'}^n)/(\Gamma^n\rtimes S_n/\Gamma_{g'}^n)=
\Sigma_{g'}^n/G_n$, with the finite group $G_n=G^n \rtimes S_n$. 
\endproof

\section{Group algebras and $K$-theory}\label{KtheorySec}

We now compute the orbifold K-theory groups of the symmetric products $\Sym^n(\Sigma)$
and we discuss their relation to the K-theory of group $C^*$-algebras.

\medskip
\subsection{Orbifold $K$-theory}

Let $X$ be a good orbifold that is orbifold covered by a smooth manifold $Y$
with $X=Y/G$. Then the orbifold $K$-theory of $X$ is given by
\begin{equation}\label{KorbX}
K^\bullet_{orb}(X)= K_\bullet(C_0(Y)\rtimes G) = K^\bullet_G(Y),
\end{equation}
the $G$-equivariant $K$-theory of $Y$.

\smallskip

We know from \cite{Farsi}, \cite{MaMa1} that, for a good $2$-dimensional
orbifold $\Sigma=\Sigma(g,\underline{\nu})$, with $m$ cone points $x_j$ with stabilizers of
order $\nu_j$, the orbifold $K$-theory is given by
\begin{equation}\label{KorbSigma}
K^\bullet_{orb}(\Sigma)=\left\{ \begin{array}{ll} \Z^{2-m+\nu} & \bullet =0 \\
\Z^{2g} & \bullet =1 \end{array}\right. ,
\end{equation}
where $\nu=\sum_{j=1}^m \nu_j$.

\smallskip

For the symmetric products $\Sym^n(\Sigma)$, using the covering (4) of Proposition
\ref{orbcoverings}, we obtain
\begin{equation}\label{orbKsymn}
K^\bullet_{orb}(\Sym^n(\Sigma))= K_\bullet ( (C(\Sigma_{g'})\rtimes G)^{\otimes n} \rtimes S_n) =
K^\bullet_{G^n\rtimes S_n}(\Sigma_{g'}^n), 
\end{equation}
where $K_\bullet(C(\Sigma_{g'})\rtimes G))=K^\bullet_{orb}(\Sigma)$. Using the
orbifold cover (1) of Proposition \ref{orbcoverings}, we see that it can also be described as
\begin{equation}\label{orbKsymn2}
K^\bullet_{orb}(\Sym^n(\Sigma))=K^\bullet_{\Gamma^n\rtimes S_n}(\H^n).
\end{equation}

\smallskip

These descriptions of the orbifold $K$-theory of the symmetric products
fall into a general framework for studying equivariant $K$-theory
with respect to the action on powers $X^n$ of the wreath products
\begin{equation}\label{wreath}
G \sim S_n := G^n \rtimes S_n ,
\end{equation}
for a finite group $G$ acting on a smooth manifold $X$.
Several important properties of the equivariant $K$-theory groups
$K^\bullet_{G^n\rtimes S_n}(X^n)$ where studied in
\cite{Wang}. We recall some of the main results of \cite{Wang}
and we apply them to our case, described as in \eqref{orbKsymn}.

\medskip
\subsection{Classifying space for proper action and assembly map}

It is known from \cite{BCH} that, to a locally compact
group $G$ one can associate a universal space for proper actions
$\underline{E}G$, and a classifying space for proper actions given
by the quotient $\underline{B}G=\underline{E}G/G$, so that there
is a Kasparov assembly map from the
equivariant $K$-homology groups with $G$-compact support
$K^G_\bullet(\underline{E}G)$ to the $K$-theory of the reduced
group $C^*$-algebra
\begin{equation}\label{muKasp}
\mu: K^G_j(\underline{E}G)\to K_j(C^*_r(G)),
\end{equation}
which assigns to an abstract $G$-equivariant elliptic operator its index.
The group $G$ satisfies the Baum--Connes conjecture if the map
\eqref{muKasp} is an isomorphism.

\smallskip

The Baum--Connes conjecture (in fact the stronger form with coefficients)
is implied by the Haagerup property, \cite{Mis}. All finite groups satisfy
the Haagerup property, and Fuchsian groups are also in the list of groups
that are known to satisfy it, see \cite{CCJJV}, \cite{Mis}. However, while
it is known from \cite{CoStaVa} that the class of groups satisfying the
Haagerup property is closed under wreath products, this only refers to
``standard" wreath products $G\sim H:=G^{(H)} \rtimes H$ where 
$G^{(H)}=\oplus_{h\in H} G$. A more general class of wreath products,
which includes the case $G^n \rtimes S_n$ that we are interested in, is
given by the ``permutation wreath products" $G\sim_X H =G^{(X)}\rtimes H$, 
where $X$ is an $H$-set and the action of $H$ on $G^{(X)}=\oplus_{x\in X} G$
is by permuting indices $x\in X$ with the $H$ action.
As shown in \cite{CoStaVa}, the Haagerup property is a lot more delicate
for the case of the permutation wreath products.

\smallskip

However, for two groups $G$ and $H$ that both satisfy the Haagerup property,
even if the more general permutation wreath products 
$G\sim_X H =G^{(X)}\rtimes H$ do not necessarily satisfy the Haagerup
property, they all do satisfy the Baum--Connes conjecture. This follows
from the general result of Oyono--Oyono on Baum--Connes for
certain group extensions, \cite{Oyo}. Thus, we have the following
property.

\begin{lem}\label{BCwreath}
The groups $\Gamma^n\rtimes S_n$, with $\Gamma=\Gamma(g,\underline{\nu})$
a Fuchsian group, satisfy the Baum--Connes conjecture, hence the assembly map
\begin{equation}\label{mumapGn}
\mu: K^\bullet_{orb}(\Sym^n(\Sigma))=K^\bullet_{\Gamma^n\rtimes S_n}(\H^n) \to K_\bullet(C^*_r(\Gamma^n\rtimes S_n))
\end{equation}
is an isomorphism.
\end{lem}

\proof The general result of \cite{Oyo} implies that the wreath products $\Gamma^n\rtimes S_n$
satisfy the Baum--Connes conjecture. The result then follows by identifying 
$\H^n=\underline{E} (\Gamma^n\rtimes S_n)$ and $\Sym^n(\Sigma)=\underline{B} (\Gamma^n\rtimes S_n)$,
with models for the universal and classifying space for proper actions, respectively.
To see this, we can use the fact that if for a group $G$ a $G$-space $Y$ is a model of the universal
space for proper actions $\underline{E}G$, and $G'\subset G$ is a subgroup, then $Y$ is also a model
of $\underline{E}G'$ (Corollary 1.9 of \cite{BCH}) and that if $G$ is a Lie group and $K$ is the maximal
compact subgroup, then a model of $\underline{E}G$ is given by the quotient $G/K$. 
A Lie group is virtually connected if it has only finitely many connected components. For
any virtually connected Lie group the quotient $G/K$ is diffeomorphic to a Euclidean space. 
We apply the above to the group $\PSL(2,\R)^n \rtimes S_n$.
\endproof

\medskip
\subsection{Orbifold $K$-theory and Lie group quotient}

Consider as above the virtually connected Lie group $\PSL(2,\R)^n \rtimes S_n$ and let $\cK_n$
denote its maximal compact subgroup, with quotient $\PSL(2,\R)^n \rtimes S_n/\cK_n \simeq \H^n$.
The orbifold symmetric product is obtained as the double quotient
$$ \Sym^n(\Sigma) = \Gamma_n \backslash \PSL(2,\R)^n \rtimes S_n/\cK_n, $$
where $\Gamma_n=\Gamma^n\rtimes S_n$. Let
\begin{equation}\label{Pnquot}
 \cP_n :=  \Gamma_n \backslash \PSL(2,\R)^n \rtimes S_n ,
\end{equation} 
\begin{equation}\label{hatPnquot}
 \hat\cP_n :=  \Gamma_{g'}^n \backslash \PSL(2,\R)^n \rtimes S_n .
\end{equation} 
Then we have the following.

\begin{lem}\label{MorEqPhatP}
Let $\cP_n$, $\hat\cP_n$, and $\cK_n$ be as above and let $G_n=G^n\rtimes S_n$, with
$G=\Gamma/\Gamma_{g'}$. 
The algebras $C_0(\cP_n)\rtimes \cK_n$ and $C_0(\hat\cP_n/\cK_n) \rtimes G_n=C(\Sigma_{g'}^n)\rtimes G_n$ are
strongly Morita equivalent.
\end{lem}

\proof By \eqref{Pnquot} and \eqref{hatPnquot}, and the fact that $\Gamma_{g'}^n \subset \Gamma^n\rtimes S_n$
is a normal subgroup  with quotient $G_n=G^n\rtimes S_n$, we obtain
$$  \cP_n = G_n \backslash \hat\cP_n , \ \ \ \text{ and } \ \ \  \hat\cP_n/\cK_n = \Gamma_{g'}^n\backslash 
\H^n =\Sigma_{g'}^n. $$
The Morita equivalence then follows as in Proposition 1.2 of \cite{MaMa1}, by applying \cite{Green}.
\endproof

The orbifold $C^*$-algebra is defined in \cite{Farsi} as
$$ C^*(\Sym^n(\Sigma)) = C(\cF_n) \rtimes SO(2n), $$
where $\cF_n$ is the frame bundle of the orbifold tangent bundle of $\Sym^n(\Sigma)$. By the same
argument of \cite{Farsi} it is shown to be strongly Morita equivalent to
$$ C^*(\Sym^n(\Sigma)) \simeq C(\Sigma_{g'}^n) \rtimes G_n. $$
Thus, combining \cite{Farsi} with Proposition \ref{orbcoverings}, we obtain the following.

\begin{cor}\label{allMoreqs}
The algebras $C^*(\Sym^n(\Sigma))$, $C(\Sigma_{g'}^n)\rtimes G_n$, $C_0(\H^n)\rtimes \Gamma_n$,
$C_0(\bS^n(\H))\rtimes \Gamma^{ab}$, 
and $C_0(\cP_n)\rtimes \cK_n$ are all strongly Morita equivalent.
\end{cor}

\medskip
\subsection{A notion of orbifold-Jacobian}

Given a good $2$-dimensional orbifold $\Sigma=\Sigma(g,\underline{\nu})$, we define the
orbifold-Jacobian of $\Sigma$ to be the product
\begin{equation}\label{orbJac}
J^{orb}(\Sigma):=  J(\Sigma) \times \prod_{j=1}^m \mu_{\nu_j} 
\end{equation}
where $J(\Sigma)=H^1(\Sigma,\R)/H^1(\Sigma,\Z)=\bT^{2g}$, a real torus of
rank $2g$, and $\mu_{\nu_j}$ denotes the group of roots of unity of order $\nu_j$.
The group structure on $J^{orb}(\Sigma)$ is the direct product
$\bT^{2g} \times \prod_{j=1}^m \mu_{\nu_j}$.

\smallskip

For each cone point $x_j$, $j=1,\ldots,m$ on the orbifold $\Sigma$, let $C_j$ be
the boundary of a small disc in $\Sigma$ centered at $x_j$. Let $x_{j,k}$ for
$k=1,\ldots, \nu_j$ be points on $C_j$. For such a collection of base points,
we define an orbifold-Abel-Jacobi map $\cA^{orb}=\{ \cA_{j,k} \}$ with
$$ \cA_{j,k}: \Sigma \to \bT^{2g} \times \{ \zeta_{j,k} \}, \ \ \ 
\cA_{j,k}: \omega \mapsto \int_{x_{j,k}}^x \omega $$
where $\zeta_{j,k}$ are the roots of unity in $\mu_{\nu_j}$. This extends to
an orbifold-Abel-Jacobi map $\cA^{orb}: \Sym^n(\Sigma) \to J^{orb}(\Sigma)$ by
$\cA^{orb}[x_1,\ldots, x_n]=\cA^{orb}(x_1)+\cdots + \cA^{orb}(x_n)$.

\medskip
\subsection{$K$-theory and the orbifold-Jacobian}

The reduced group $C^*$-algebra $C^*_r(G)$ of a discrete group $G$ is 
the norm closure in the algebra of bounded operators on $\ell^2(G)$ of the 
group ring $\C[G]$, acting via the left regular representation
$L_g \xi (g') =  \xi(g^{-1} g')$.

\begin{lem}\label{Kthlem}
The reduced group $C^*$-algebra $C^*_r(\pi_1^{orb}(\Sym^n(\Sigma)))$
has $K$-theory isomorphic to the topological $K$-theory of the orbifold-Jacobian
$J^{orb}(\Sigma)$.
\end{lem}

\proof
The group $\pi_1^{orb}(\Sym^n(\Sigma))=\pi_1^{orb}(\Sigma)^{ab}$ is abelian.
Thus, the $K$-theory of the reduced group $C^*$-algebra can be identified with
the topological $K$-theory of the dual group, under Pontrjagin duality,
\begin{equation}\label{Kab}
K_j (C^*_r(\pi_1^{orb}(\Sigma)^{ab}))\simeq K_j(C(\widehat{\pi_1^{orb}(\Sigma)^{ab}})).
\end{equation}
We have $\pi_1^{orb}(\Sigma)^{ab}=\Z^{2g}\oplus \bigoplus_j \Z/\nu_j \Z$.
The Pontrjagin dual of a direct sum of abelian groups is the direct product
of the Pontrjagin duals. The dual of $\Z^{2g}$ is a $2g$-dimensional real torus
$\bT^{2g}=S^1\times \cdots \times S^1$, while for each finite group $\Z/\nu_j\Z$
the Pontrjagin dual is the subgroup $\mu_{\nu_j} \subset S^1$ of $\nu_j$-th roots 
of unity, which can be identified again with $\Z/\nu_j\Z$. Thus, we obtain a direct product
$$ \widehat{\pi_1^{orb}(\Sigma)^{ab}} = \bT^{2g} \times \prod_j \mu_{\nu_j}. $$
\endproof

\medskip
\subsection{A homotopy theoretic version}

Consider a smooth surface $\Sigma_{g'}=\H/\Gamma_{g'}$ uniformized
by the hyperbolic plane $\H$. The surface $\Sigma_{g'}$ is a model of
the classifying space for proper actions $\Sigma_{g'}=\underline{B}\Gamma_{g'}$,
with $\H=\underline{E}\Gamma_{g'}$ a model for a universal space for proper
actions, \cite{BCH}. 

\smallskip

It is known (see Theorem 1.1 of \cite{Val}) that a group homomorphism
$\alpha: \Gamma_1 \to \Gamma_2$ induces a commutative diagram
\begin{equation}\label{Kdiagram}
 \xymatrix{ K_\bullet^{\Gamma_1}(\underline{E}\Gamma_1)\dto^{\alpha_*} \rto^{\mu} &
K_\bullet (C^*(\Gamma_1)) \dto^{\alpha_*} \\
K_\bullet^{\Gamma_2}(\underline{E}\Gamma_2) \rto^{\mu} &
K_\bullet (C^*(\Gamma_2)) .
} \end{equation}
The analogous statement for reduced algebras $C^*_r(\Gamma_i)$ holds in
general only for monomorphisms (Corollary 1.2 of \cite{Val}), but since the geometric
left-hand-side is always functorial, under the hypothesis that the Baum--Connes
conjecture holds, then the right-hand-side would also be functorial for the reduced case,
as observed in \cite{Val}. We focus on the case where $\Gamma_1=\Gamma_{g'}$
and $\Gamma_2=\Gamma_{g'}^{ab}=H^1(\Sigma_{g'},\Z)$, with 
$\alpha: \Gamma_{g'} \to \Gamma_{g'}^{ab}$ the quotient map. In this case,
we know that the groups involved satisfy the Baum--Connes conjecture, and
we can think of the left-hand-side of the diagram \eqref{Kdiagram} as a kind of 
``homotopy-theoretic Abel-Jacobi map" from the $K$-homology of the curve 
$\Sigma_{g'}$ to that if its Jacobian, 
\begin{equation}\label{Kdiagram2}
 \xymatrix{ K_\bullet^{\Gamma_{g'}}(\H) \simeq  K_\bullet(\Sigma_{g'}) 
 \dto^{\alpha_*} \rto^{\ \mu} &
K_\bullet (C^*(\Gamma_{g'})) \dto^{\alpha_*} \\
K_\bullet^{\Gamma_{g'}^{ab}}(\R^{2g'}) \simeq  K_\bullet(J(\Sigma_{g'})) \rto^{\ \ \ \mu} &
K_\bullet (C^*(\Z^{2g'})) .
} \end{equation}
In a similar way, we obtain maps
\begin{equation}\label{OrbKdiag}
K^{\Gamma}_\bullet(\H) \simeq K^{orb}_\bullet(\Sigma) \stackrel{\mu}{\to} K_\bullet(C^*(\Gamma))
\to K_\bullet(C^*(\Gamma^{ab}))=K^\bullet(J^{orb}(\Sigma))
\end{equation}
and similar maps for the symmetric products
\begin{equation}\label{OrbKdiag2}
K^{\Gamma^n\rtimes S_n}_\bullet(\H^n) \simeq K^{orb}_\bullet(\Sym^n(\Sigma)) 
\stackrel{\mu}{\to} K_\bullet(C^*(\Gamma^n\rtimes S_n))
\to K_\bullet(C^*(\Gamma^{ab}))=K^\bullet(J^{orb}(\Sigma)),
\end{equation}
where the last map is induced by the quotient
map $\Gamma^n\rtimes S_n \to \Gamma^n\rtimes S_n/\langle S_n \rangle \simeq \Gamma^{ab}$.

\medskip

\section{The magnetic field and twisted group algebras}\label{magfieldSec}

\medskip
\subsection{Twisted group ring and twisted group $C^*$-algebra}\label{TwistDefSec}
Recall that, for a discrete group $\Gamma$, a multiplier on $\Gamma$ is defined as a map
$\sigma: \Gamma \times \Gamma\to U(1)$ satisfying the properties:
\begin{enumerate}
\item $\sigma(\gamma,1)=\sigma(1,\gamma)=1$, for all $\gamma\in \Gamma$,
\item $\sigma(\gamma_1,\gamma_2)\sigma(\gamma_1\gamma_2,\gamma_3) = \sigma(\gamma_1,\gamma_2\gamma_3)\sigma(\gamma_2,\gamma_3)$, for all $\gamma_1,\gamma_2,\gamma_3\in \Gamma$.
\end{enumerate}
The reduced twisted group $C^*$-algebra $C^*_r(\Gamma,\sigma)$ is the norm closure in the algebra of bounded
operators on $\ell^2(\Gamma)$ of the twisted group ring $\C[\Gamma,\sigma]$, generated by the left translations
$L^\sigma_\gamma$ with relations $L^\sigma_\gamma L^\sigma_{\gamma'}=\sigma(\gamma,\gamma') L^\sigma_{\gamma\gamma'}$,
represented on $\ell^2(\Gamma)$ by the left regular representation
\begin{equation}\label{leftregrep}
L^\sigma_\gamma \xi (\gamma') = \sigma(\gamma, \gamma^{-1}\gamma') \, \xi(\gamma^{-1} \gamma').
\end{equation}

\medskip

\subsection{Cocycles from the magnetic field}

In the system we are considering, we have $n$ indistinguishable particles
moving in a negatively curved geometry $\H$, subject to a potential, which is generated by
charges disposed along the vertices of an embedding (Cayley graph) of a Fuchsian group
$\Gamma$ in $\H$, and to an external magnetic field. 

\smallskip

Usually, in the ``independent electron approximation" in the theory of solids, one 
replaces the (unbounded) interaction potential of the many-particle problem with
a Hamiltonian with an effective potential of the form
$H= \sum_{i=1}^n -\Delta_{x_i} + V(x_i)$,  with a (bounded) effective potential
$V(x)$ of a single-particle problem that encodes an average of the interactions
of one of the charge carriers with the others, as well as with the atoms of the
periodic medium. Here we consider a more general situation, where the independent 
electron approximation is perturbed by a (bounded) periodic interaction potential. This
means that we consider a Hamiltonian as above with the $V(x_i)$  replaced by
a smooth bounded function $W(x_1,\ldots,x_n)$ on $\H^n$, which is invariant 
under the symmetry group $\Gamma^n\rtimes S_n$.

\medskip

As in \cite{CHMM}, \cite{MaMa1}, \cite{MaMa2}, \cite{MaMa3} the magnetic field
is described by a closed $2$-form $\omega=d\eta$ on $\H$. The form is invariant under
the action of $\Gamma$ on $\H$, so that $\omega -\gamma^*\omega =0$ for all
$\gamma\in \Gamma$. However, the potential $\eta$ is not $\Gamma$-invariant.
The fact that $d(\eta -\gamma^*\eta)=0$ for all $\gamma\in \Gamma$ implies the
existence of a (real valued) $0$-form $\phi_\gamma$ on $\H$ satisfying
$\gamma^*\eta -\eta = d \phi_\gamma$, for all $\gamma\in \Gamma$. The
function 
\begin{equation}\label{phigamma}
\phi_\gamma =\int_{x_0}^x \gamma^*\eta-\eta
\end{equation} 
satisfies the property that 
\begin{equation}\label{phigammaindep}
\phi_\gamma(x)
+\phi_{\gamma'}(\gamma x)-\phi_{\gamma'\gamma}(x) = \phi_{\gamma'}(\gamma x_0) 
\end{equation}
is independent of $x\in \H$. Setting 
\begin{equation}\label{sigmaphi}
\sigma(\gamma,\gamma')= \exp(-i \phi_{\gamma'}(\gamma x_0)),
\end{equation}
for a chosen base point $x_0\in \H$, determines a multiplier of $\Gamma$, as in
\S \ref{TwistDefSec}. The magnetic Laplacian $\Delta^\eta = (d-i\eta)^* (d-i\eta)$ is
invariant under the magnetic translations $T^\phi_\gamma = e^{-i\phi_\gamma} T_\gamma$
and the algebra of magnetic translations satisfies
\begin{equation}\label{magtransprod}
T^\phi_\gamma\, T^\phi_{\gamma'} =\sigma(\gamma,\gamma')\, T^\phi_{\gamma\gamma'},
\end{equation}
with $\sigma(\gamma,\gamma')$ as in \eqref{sigmaphi}. This follows from \eqref{phigammaindep}.

\smallskip

Consider now the product $\H^n$ and the $2$-form 
$\varpi=\sum_j \omega_j$, where $\omega_j$ is
the pullback $\omega_j =\pi_j^* \omega$ of the magnetic field $2$-form
described above, under the projection of $\H^n$ onto the $j$-th factor.
In particular $\omega_j$ only depends on the $j$-coordinate of $\H^n$.

\begin{lem}\label{2nformsigma}
The $2$-form $\varpi$ on $\H^n$ is invariant under the action of $\Gamma_n=\Gamma^n\rtimes S_n$.
The potential, given by the $1$-form $\zeta =\sum_j \eta_j$ is invariant under $S_n$, 
hence it descends to a $1$-form on $\Sym^n(\H)=\H^n/S_n$.
The form $\zeta$, moreover, satisfies $g^*\zeta - \zeta = d\psi_g$, for
$g=(\gamma,\sigma)\in \Gamma^n\rtimes S_n$, with 
\begin{equation}\label{psigamma}
\psi_g(x)=\sum_{j=1}^n \phi_{\gamma_j}(x_{\sigma(j)}),
\end{equation}
with $\phi_{\gamma_j}$ as in \eqref{phigamma} with a base point $x_{(0)}=( x_{0,j} )$. 
This function $\psi_g : \H^n \to \R$ satisfies
\begin{equation}\label{psiindep}
\psi_g(x) + \psi_{g'}(gx) - \psi_{g'g}(x) = \psi_{g'}(g x_{(0)}),
\end{equation}
independent of $x=(x_j)\in \H^n$. This determines a multiplier $\sigma_n: \Gamma_n \times \Gamma_n \to U(1)$,
\begin{equation}\label{sigman}
\sigma_n (g,g')= \exp(-i \phi_{g'}(g \, x_{(0)})).
\end{equation}
For $g\in \Gamma_n$, the magnetic translations $T^\psi_g$ on $\H^n$ 
satisfy
\begin{equation}\label{magtranslHn}
T^\psi_{g'} T^\psi_{g} = \sigma_n (g,g') T^\psi_{g'g} .
\end{equation}
\end{lem}

\proof The identity \eqref{phigammaindep} implies that for all $j=1,\ldots, n$ and all $\gamma_j\in \Gamma$,
and all $\sigma, \sigma'\in S_n$ we have
$$ \phi_{\gamma_{\sigma'(j)}}(x_{\sigma'\sigma(j)}) + \phi_{\gamma'_j}(\gamma_{\sigma'(j)}x_{\sigma'\sigma(j)})
- \phi_{\gamma'_j \gamma_{\sigma'(j)}}(x_{\sigma'\sigma(j)})=
\phi_{\gamma'_j}(\gamma_{\sigma'(j)} x_{0,\sigma'\sigma(j)}), $$
so that by summing over $j$ we obtain \eqref{psiindep}. The 
composition of two magnetic translations then gives
$$ e^{-i \psi_{g}(x)} e^{-i \psi_{g'}(g x)} f(g'g x) = e^{i(\psi_{g}(x)+\psi_{g'}(g x) 
- \psi_{g'g}(x))} T^\psi_{g'g}\, f(x), $$
so we obtain \eqref{magtranslHn} with the multiplier \eqref{sigman}.
\endproof

\medskip
\subsection{$K$-theory of the twisted group algebra}

The multiplier $\sigma_n: \Gamma_n \times \Gamma_n \to U(1)$ described above 
determines a twisting $C^*_r(\Gamma_n,\sigma_n)$
of the group $C^*$-algebra. The twisted group algebra provides the 
algebra of observables for the $n$-particle system, in the presence of the external magnetic field.

\smallskip

In \cite{MaMa1} it was shown that the $K$-theory of the twisted group algebra
$C^*_r(\Gamma,\sigma)$ is isomorphic to the $K$-theory of the untwisted algebra,
whenever the multiplier $\sigma: \Gamma \times \Gamma \to U(1)$ has trivial 
Dixmier--Douady class $\delta(\sigma)=0$.
In the case of the algebras $C^*_r(\Gamma_n,\sigma_n)$, we have a similar results,
as we will now discuss. 

\medskip

\begin{prop}\label{sigmasigman}
Let $\sigma : \Gamma \times \Gamma \to U(1)$ be a multiplier, with trivial 
Dixmier--Douady class, $\delta[\sigma]=0$. Then it determines a multiplier
$\sigma_n: \Gamma_n \times \Gamma_n \to U(1)$, which also has
trivial Dixmier--Douady invariants, $\delta[\sigma_n]=0$. 
\end{prop}

\proof Recall that the exponential sequence
$$ 1 \to \Z \stackrel{\iota}{\to} \R \stackrel{e}{\to} U(1) \to 1, $$
with $e(t)=\exp(2\pi i t)$, determines a long exact cohomology sequence
$$ \cdots \to H^2(\Gamma, \Z) \stackrel{\iota_*}{\longrightarrow} H^2(\Gamma,\R) \stackrel{e_*}{\longrightarrow} 
H^2(\Gamma, U(1))
\stackrel{\delta}{\longrightarrow} H^3(\Gamma,\Z) \stackrel{\iota_*}{\longrightarrow} H^3(\Gamma,\R) \to \cdots $$
where $\delta: H^2(\Gamma, U(1))\to H^3(\Gamma,\Z)$ is the Dixmier--Douady map. 
The Fuchsian group $\Gamma=\Gamma(g,\underline{\nu})$ has cohomology (see \cite{MaMa1})
\begin{equation}\label{HjGammaR}
H^j(\Gamma,\R) = \left\{ \begin{array}{ll} \R & j=0,2 \\
\R^{2g} & j=1 \\
0 & j\geq 3
\end{array}\right. 
\end{equation}
A multiplier $\sigma:\Gamma \times \Gamma \to U(1)$ determines a cocycle $\sigma \in Z^2(\Gamma,U(1))$,
with cohomology class $[\sigma]\in H^2(\Gamma,U(1))$. If the Dixmier--Douady class $\delta[\sigma]=0$
in $H^3(\Gamma,\Z)$, then the class $[\sigma]$ is in the range of the map $e_*: H^2(\Gamma,\R) \to H^2(\Gamma, U(1))$,
that is, there exists a cocycle $\xi \in Z^2(\Gamma,\R)$ such that $[\sigma]=[e(\xi)]$. Using the branched covering of
the $2$-dimensional orbifold $\Sigma$ by a smooth Riemann surface $\Sigma_{g'}$ with
$\Sigma =\Sigma_{g'}/G$, for a finite group $G$, we can identify $H^2(\Gamma,\R)\cong H^2(\Sigma_{g'},\R)$,
since the finite group $G$ has no nontrival cohomology with real coefficients. Thus, we can realize the cocyle
$\xi$ in terms of a closed $2$-form $\omega$ on $\Sigma_{g'}$, or of its $\Gamma_{g'}$-invariant lift to the
universal cover $\H$, with $[\sigma]=[e(\omega)]$. 
Consider then the cohomology $H^2(\Gamma_n, \R)$. By the results of \cite{Leary} \cite{Naka}
we know that the Lyndon--Hochschild--Serre spectral sequence for the 
group cohomology of the wreath product $\Gamma^n\rtimes S_n$ degenerates at the 
$E_2$-term,  for both integral cohomology and cohomology with coefficients in a field.
In particular, this means that we can compute the cohomology of $\Gamma_n$ with real
coefficients in terms of the cohomology groups $E^{pq}_2=H^p(S_n, H^q(\Gamma^n,\R))$, with
$H^q(\Gamma^n,\R)=\oplus_{i_1+\cdots +i_n=q} H^{i_1}(\Gamma,\R)\times \cdots \times H^{i_n}(\Gamma,\R)$. For  $p+q=2$, the only non-trivial term is $H^0(S_n, H^2(\Gamma^n,\R))$, 
since for the symmetric group $H^j(S_n,\R)=0$ for $j\geq 1$. 
There is a subspace in $H^0(S_n, H^2(\Gamma^n,\R))\cong H^2(\Gamma^n,\R)$
that is isomorphic to $H^2(\Gamma,\R)^{\oplus n}$, namely the subspace given by the 
K\"unneth components involving only $H^2$ and $H^0$ and not $H^1$. This
subspace can be identified with $H^2(\Sigma_{g'},\R)^{\oplus n}$, as above.
Using this identification,
we see that the closed $2$-form $\varpi(x) =\sum_j \omega(x_j)$ on $\Sigma_{g'}^n$
determines a class $[\varpi]$ in this subspace of $H^2(\Gamma_n,\R)$, hence it determines
a multiplier class $[\sigma_n]=[e(\varpi)] \in H^2(\Gamma_n,U(1))$, as the image 
under the map $e_*$ in the cohomology exact sequence
\begin{equation}\label{Gammanexactseq}
 \cdots \to H^2(\Gamma_n, \Z) \stackrel{\iota_*}{\longrightarrow} H^2(\Gamma_n,\R) \stackrel{e_*}{\longrightarrow} 
H^2(\Gamma_n, U(1))
\stackrel{\delta}{\longrightarrow} H^3(\Gamma_n,\Z) \stackrel{\iota_*}{\longrightarrow} H^3(\Gamma_n,\R) \to \cdots 
\end{equation}
By construction, a representative $\sigma_n: \Gamma_n \times \Gamma_n \to U(1)$ of
this class will be a multiplier with trivial Dixmier--Douady class, $\delta [\sigma_n]=0$.
\endproof

\smallskip

\begin{rem}\label{omegarem}{\rm
In the case of the multiplier $\sigma: \Gamma \times \Gamma \to U(1)$ defined by the magnetic field, 
we can take, in the argument of Lemma \ref{sigmasigman}, 
the $\Gamma$-invariant $2$-form $\omega$ on $\H$ given by the magnetic field.
The corresponding multiplier $\sigma_n: \Gamma_n \times \Gamma_n \to U(1)$ will then agree
with the one constructed in the previous subsection.
Thus, the $\sigma$ and the $\sigma_n$ determined by the magnetic field have trivial 
Dixmier--Douady invariant.  }
\end{rem}

\medskip

With the $K$-amenability property discussed in \S \ref{Kamen} below, we
have the following.

\begin{prop}\label{Ksigma}
Let  $\sigma_n: \Gamma_n \times \Gamma_n \to U(1)$ be a multiplier as above, with $\delta[\sigma_n]=0$. 
Then $K_\bullet(C^*_r(\Gamma_n,\sigma_n))\simeq K_\bullet(C^*_r(\Gamma_n))$.
\end{prop}

\proof The argument is the same as in \cite{MaMa1}.
If $\delta[\sigma_n]=0$, we have $[\sigma_n]=[e(\xi_n)]$ for some $\xi_n \in Z^2(\Gamma_n, \R)$,
by \eqref{Gammanexactseq}. We can then use a homotopy $[\sigma_{n,t}]=[e(t \xi_n)]$ with $0\leq t \leq 1$.
Consider the discrete subgroup $\Gamma_n$ of $\cG_n=\PSL(2,\R)^n\rtimes S_n$,
with quotient $\Gamma_n \backslash \cG_n = \cP_n$ and
let $\cA$ be an algebra with an action of $\Gamma_n$ by automorphisms. The
crossed product $(\cA\otimes C_0(\cG_n)) \rtimes \Gamma_n$ is Morita equivalent to 
the algebra of sections $C_0(\Gamma_n \backslash \cG_n, \cE)$ of the flat $\cA$-bundle
$\cE \to \hat\cP_n$ with $\cE=(\cA \times \cG_n)/\Gamma_n$ with the quotient taken with
respect to the diagonal action. Moreover, the algebras $(\cA\rtimes \Gamma_n) \otimes C_0(\cG_n)$
and $(\cA\otimes C_0(\cG_n))\rtimes \Gamma_n$ have the same $\cK_n$-equivariant $K$-theory.
Combined with the previous Morita equivalence and the fact that $\cG_n/\cK_n=\H^n$, we obtain
$$ K_{\cK_n,\bullet} (C_0(\cP_n,\cE)) \cong K_{\cK_n,\bullet+\dim(\cG_n/\cK_n)}(\cA\rtimes \Gamma_n)
=K_{\cK_n,\bullet}(\cA\rtimes \Gamma_n). $$
As in \cite{MaMa1}, we use the Packer--Raeburn stabilization trick \cite{PaRae}.
The algebra 
$\cA\rtimes_{\sigma_n}\Gamma_n$ is stably isomorphic to $(\cA\otimes \bK)\rtimes \Gamma$ with
$\bK$ the algebra of compact operators, and we consider the flat $\cA\otimes \bK$-bundle
$$ \cE_{\sigma_n} =(\cA\otimes \bK\times \cG_n)/\Gamma_n \to \Gamma_n \backslash \cG_n. $$
As in Proposition 2.2 of \cite{MaMa1}, we then have
$$ K_\bullet(C^*(\Gamma_n, \sigma_n)) \cong K_{\cK}^{\bullet} (\Gamma_n\backslash \cG_n, \delta(B_{\sigma_n})). $$
Here the twisted $\cK_n$-equivariant $K$-theory $K_{\cK}^{\bullet} (\Gamma_n\backslash \cG_n, \delta(B_{\sigma_n}))$
is the same as the $\cK_n$-equivariant $K$-theory of the continuous trace $C^*$-algebra 
$B_{\sigma_n}=C_0(\Gamma_n\backslash \cG_n,\cE_{\sigma_n})$ 
with Dixmier--Douady class $\delta(B_{\sigma_n})$. By Theorem 2.3 of \cite{MaMa1}, using the
$K$-amenability property of \S \ref{Kamen} below, 
we then have $K_\bullet(C^*(\Gamma_n,\sigma_n))\cong K_\bullet(C^*_r(\Gamma_n,\sigma_n))$.
We then obtain isomorphisms $K_\bullet(C^*_r(\Gamma_n, \sigma_{n,t}))=K_\bullet(C^*_r(\Gamma_n))$.
\endproof

Combining the isomorphism $K_\bullet(C^*_r(\Gamma_n, \sigma_{n,t}))=
K_\bullet(C^*_r(\Gamma_n))$ obtained above with the Kasparov assembly map, we
obtain a {\em twisted Kasparov map} (as in \cite{MaMa1}, \cite{MaMa2})
\begin{equation}\label{twistKasp}
\mu_{\sigma_n} : K^\bullet_{rob}(\Sym^n(\Sigma)) \to K_\bullet(C^*_r(\Gamma_n, \sigma_n)).
\end{equation}

\medskip
\subsection{K-amenability}\label{Kamen}

We now turn to the K-amenability property of the group $\SL(2,\R)^n\rtimes S_n$.
We first recall some basic facts about K-amenability.
A locally compact second countable group $\cG$ is amenable if the map
$\lambda: C^*(\cG) \to C^*_r(\cG)$ determined by restriction of representations is an
isomorphism. In particular, if $\cG$ acts on a $C^*$-algebra $A$ as a $C^*$-dynamical
system, and $\cG$ is amenable, then the map $\lambda_A: C^*(G,A) \to C^*_r(G,A)$
is also an isomorphism. The notion of K-amenability expresses a weaker K-theoretic form
of this property.  We consider here two forms of the K-amenability property, as 
in \cite{FoxHask}. Recall that a Fredholm $\cG$-module is a pair $(\cH, F)$ of
a Hilbert space $\cH=\cH_0 \oplus \cH_1$ with unitary representations $\rho_0$, $\rho_1$ 
of $\cG$ on $\cH_0$ and $\cH_1$ and with a bounded operator $F:\cH_0 \to \cH_1$ with
$$ g \mapsto \rho_1(g) \circ F- F \circ \rho_0(g) $$
a compact operator and with $F^*F-1$ and $FF^*-1$ also compact operators. The trivial
Fredholm module has $\cH_0=\C$ and $\cH_1=0$. The notion of a homotopy
of Fredholm module is similarly stated (see e.g.~\cite{FoxHask}).

\begin{enumerate}
\item $\cG$ is K-amenable if, for any $C^*$-dynamical
system $(\cG,A)$, the map $\lambda_{A\, *}: K_*(C^*(G,A)) \to K_*(C^*_r(G,A))$
is an isomorphism.
\item $\cG$ is K-amenable if there exists a Fredholm $\cG$-module $(\cH,F)$ such that
the representation of $\cG$ on the Hilbert space $\cH$ is weakly contained in the
left regular representation, with $(\cH,F)$ homotopic to the trivial Fredholm $\cG$-module.
\end{enumerate}

The second version of the K-amenability property implies the first: we are
going to refer to this second property as K-amenability. It was shown 
in \cite{Kas} that any covering group of the identity component of $SO(n,1)$ is
K-amenable. In particular, $\SL(2,\R)$ is K-amenable.  
It is also shown in \cite{Cuntz} that the class of K-amenable groups is closed with
respect to the operations of taking subgroups and taking direct products. Thus,
the groups $\SL(2,\R)^n$ are K-amenable.

\begin{prop}\label{quesKamen}
The wreath product groups $\SL(2,\R)^n\rtimes S_n$ are K-amenable.
\end{prop}

\proof The K-amenability of $\SL(2,\R)$  follows from the general
result of \cite{Kas} mentioned above. 
A more explicit proof was given in \cite{FoxHask}, by constructing a Fredholm module
with the desired properties. This has $\cH_0=L^2(K/M)$ where $K$ is the maximal
compact (the circle group), with basis $\phi_n(\theta)=e^{in\theta}$, $n\in 2\Z$.
and $M=\{\pm 1\}\subset K$, and $\cH_1=\cH_{+2} \oplus \cH_{-2}$
consisting of Hilbert space completions of the two discrete series representations, 
respectively given by the spans of $\{ w_n\,|\, n\in 2\Z, \, n\geq 2\}$ and
$\{ w_n\,|\, n\in 2\Z, \, n\leq -2\}$, and with $F(\phi_0)=0$ and $F(\phi_n)=w_n$.
See \S 1 of \cite{FoxHask} for more details.  
The K-amenability of $\SL(2,\R)^n$ follows, as mentioned above, from the general result of \cite{Cuntz}
which in particular shows the property is preserved by direct products. The construction
of a Fredholm module for $\SL(2,\R)^n$ can be obtained from the construction of 
\cite{FoxHask} by tensor products. Notice that the weak containment of representations
has the property that, if unitary representations $\pi_1$ and $\pi_2$ are, respectively,
weakly contained in unitary representations $\rho_1$ and $\rho_2$, then
$\pi_1\otimes \pi_2$ is weakly contained in $\rho_1\otimes \rho_2$.
Let $\pi$ be a unitary representation of a locally compact group $\cG$ on a Hilbert space $\cH$.
It induces a representation $\cR_\pi$ of the Banach algebra $L^1(\cG)$ on the same Hilbert space.
Moreover, one has a $\star$-homomorphism $\cR: L^1(\cG) \to C^*_r(\cG)$. The representation
$\pi$ is weakly contained in the regular representation if $\| \cR_\pi(f) \| \leq \| \cR(f) \|$ for
all $f\in L^1(\cG)$, see Definition 9.2.7 of \cite{GaDa}. For $\cG_n=\SL(2,\R)^n\rtimes S_n$, 
consider the $\cG_n$-Fredholm module given by $\cH^{\otimes n}$, with $\cH$ the Fredholm
module of \cite{FoxHask}. The operator induced by $F$ commutes with elements of $S_n$,
and one obtains in this way a $\cG_n$-Fredholm module. The representation of $\cG_n$
is still weakly contained in the regular representation, because the inequality above
is still satisfied for $f\in L^1(\cG_n)$.
\endproof

As an alternative, it may also be possible prove the result above by adapting the argument
in Proposition 2.5 and Corollary 2.6 of \cite{Cuntz}.

\section{Different notions of orbifold Euler characteristic}\label{EulerSec}

There are several different notions of orbifold Euler characteristic used in the literature.
We will recall here some of the main versions and their relation. One of the main
difference is that some orbifold Euler characteristics are rational valued, while other,
even though they appear to be defined as fractions, are in fact integer valued.
In particular, we are interested here in distinguishing between the Satake notion of
(rational valued) orbifold Euler characteristic, which plays an important role in the 
noncommutative geometry approach to the fractional quantum Hall effect developed
in \cite{MaMa1}, \cite{MaMa2}, \cite{MaMa3}, and the notion of orbifold Euler
characteristic that arises naturally in string theory, \cite{DHVW}, \cite{VaWi}.  It was
shown in \cite{HiHo}, by a simple calculation, that the latter is integer valued. We restrict
our attention here to the case of good orbifolds, which are global quotients, since the
specific cases we intend to focus on, the symmetric products $\Sym^n(\Sigma)$ of
good $2$-dimensional orbifolds, belong to this class: they are global quotients
$\Sym^n(\Sigma)=\Sigma_{g'}^n / G^n\rtimes S_n$, as we discussed above.

\smallskip

Let $X$ be a smooth manifold and $G$ a finite group, acting on $X$ with an orbifold
quotient $X/G$. Then the Satake orbifold Euler characteristic, \cite{Sat}, which we simply write
as $\chi^{orb}(X/G)$ is given by
\begin{equation}\label{chiSatake}
\chi^{orb}(X/G) = \frac{1}{\# G}\,\, \chi(X)  \,\,\, \in \Q.
\end{equation}
In \cite{Sat} a Gauss--Bonnet theorem is proved for orbifolds, where the usual topological
Euler characteristic is replaced by the orbifold version \eqref{chiSatake}, which is no
longer, in general, an integer. The index theorems for elliptic operators on orbifolds 
proved in \cite{Kawa1}, \cite{Kawa2} generalize the result of \cite{Sat}. They  
were used in \cite{MaMa1}, \cite{MaMa2} to obtain fractional values of the Hall conductance
as values of a higher twisted index theorem modeled on \cite{Kawa1} and on the higher index
theorem of \cite{CoMo}. 

\smallskip

A different notion of orbifold Euler characteristic arises naturally in the context of 
string theory on orbifolds, \cite{DHVW}, \cite{VaWi}. We will refer to it here as
``string-theoretic orbifold Euler characteristic", and we will denote it by
$\chi^{orb}(X,G)$, again assuming that the orbifold is a global quotient $X/G$
of a smooth manifold by a finite group action. This version of the 
orbifold Euler characteristic is defined as
\begin{equation}\label{chistring}
\chi^{orb}(X,G) = \frac{1}{\# G}\,\, \sum_{gh=hg} \chi(X^{\langle g, h\rangle}),
\end{equation}
where the sum is over all pairs of commuting elements in $G$ and 
$X^{\langle g, h\rangle}$ is the (common) fixed point set of $g$ and $h$. 
Although from this definition this also appears to be rational valued, it is shown
in \cite{HiHo} that the sum in \eqref{chistring} can be equivalently written as
\begin{equation}\label{chistringZ}
\chi^{orb}(X,G) = \sum_{[g]} \chi(X^g / C(g)),
\end{equation}
where now the sum is over conjugacy classes $[g]$ and $C(g)$ is the centralizer of $g$ in $G$.
The denominator $\#G$ disappears due to the simple fact that $\# [g] \cdot \# C(g) = \# G$.
In orbifold string theory, the sum in \eqref{chistring} corresponds to the sum over the different sectors. 
Notice that the Satake orbifold Euler characteristic $\chi^{orb}(X/G)$ appears in the sum 
\eqref{chistring} as the term corresponding to the trivial sector with $g=h=1$. It was shown in
\cite{AtSe} that $\chi^{orb}(X,G)={\rm rank} K^0_G(X)- {\rm rank} K^1_G(X)$, the difference
of ranks of the equivariant $K$-theory.

\smallskip

\begin{ex}\label{chiorbsSigma}{\rm 
In the case of the $2$-dimensional good orbifold $\Sigma =\Sigma_{g'}/G$ the
Satake orbifold Euler characteristic is given by $\chi^{orb}(\Sigma)=(\# G)^{-1} \chi(\Sigma_{g'})=\chi(\Sigma) +\sum_j (\nu_j^{-1}-1)$, while the string-theoretic orbifold Euler characteristic is $\chi^{orb}(\Sigma_{g'},G)=\chi(\Sigma_{g'})
+\sum_j (\nu_j-1)$.}
\end{ex}

\smallskip
\subsection{Orbifolds of $A$-sectors and inertia orbifolds}

The Satake orbifold Euler characteristic and the string-theoretic orbifold Euler characteristic
admit a family of common generalizations, see \cite{Tama}, \cite{Tama2} and
in \cite{FarSea}, \cite{FarSea2}, \cite{FarSea3}. As above, let $Y=X/G$ be a 
good orbifold.  We denote by $\cG=\cG(Y)$ the associated orbifold groupoid.
In this setting one considers the additional data of a finitely
generated discrete group $A$ and defines the orbifold $Y_A$ of $A$-sectors of $Y$
through its orbifold groupoid, which is given by $\cG(Y_A)=\cG(Y) \ltimes \Hom(A,\cG(Y))$.
One then defines the orbifold Euler characteristic $\chi^{orb}_A(X,G)$ as
\begin{equation}\label{chiAorb}
\chi^{orb}_A(X,G) = \chi^{orb}(Y_A),
\end{equation}
namely the Satake orbifold Euler characteristic of the orbifold $Y_A$.
When $A=\Z^m$ one recovers the orbifold Euler characteristics $\chi^{orb}_m(X,G)$ of
\cite{BryFul}. In particular, $\chi^{orb}_{\Z^2}(X,G)=\chi^{orb}(X,G)$, with $Y_{\Z^2}$ the inertia orbifold.

\medskip
\subsection{Orbifold Chern-Schwartz-MacPherson  classes}

A generalization of the generating function of (string theoretic) 
orbifold Euler characteristics \eqref{chiorbGenFun}
was given in \cite{Tama}, by considering the orbifold Euler characteristics 
$\chi^{orb}_A(X^n,G^n \rtimes S_n)$, with $G$ a finite group.
The case of the string-theoretic orbifold Euler characteristics of \eqref{chistring} and \eqref{chiorbGenFun} 
is recovered for $A=\Z^2$.
A further generalization of both \eqref{chiorbGenFun} and the result of \cite{Tama} 
was obtained in \cite{Ohmoto} as a generating function of orbifold characteristic classes, where the latter
are defined as (equivariant) Chern-Schwatz-MacPherson classes, whose zero-dimensional component 
recovers the Euler characteristic. The notion of orbifold Chern-Schwatz-MacPherson classes considered
in \cite{Ohmoto}  is closely related to the {\em stringy Chern classes} of \cite{Alu} and \cite{Def}.
The orbifold CSM class is defined in \cite{Ohmoto} as
the image under the equivariant MacPherson natural transformation $C_*^{G_n}$ (see \cite{Ohmoto2})
of the canonical constructible function
$$ {\bf 1}^{\Z^2}_{\Sigma_{g'}^n, G_n} = \frac{1}{\#G_n} \sum_{\rho\in \Hom(\Z^2, G_n)} {\bf 1}_{(\Sigma_{g'}^n)^{\rho(\Z^2)}}, $$
where $(\Sigma_{g'}^n)^{\rho(\Z^2)}$ is the fixed point set of the action of $\rho(\Z^2) \subset G_n$ on
$\Sigma_{g'}^n$.
Then the generating function of the orbifold CSM classes is then obtained by applying 
$C^{G,sym}_*=\prod_n C^{G_n}_*$ to the series $\sum_n {\bf 1}^{\Z^2}_{\Sigma_{g'}^n, G_n} q^n$.
This gives (Proposition 4.2 of \cite{Ohmoto})
$$ \sum_n C^{G_n}_*(\Sigma_{g'}^n) \, q^n = \prod_{\ell=1}^\infty (1-q^\ell \Delta^\ell)^{- C^G_*(\Sigma_{g'})}, $$
as in Theorem 1.2 of \cite{Ohmoto} with $A=\Z^2$, and with $\Delta^\ell$ the morphism on homology
induced by the diagonal embedding $\Delta:  \Sigma_{g'} \hookrightarrow \Sigma_{g'}^\ell$.
The string-theoretic  orbifold Euler characteristics \eqref{chiorbGenFun} are obtained by
taking the $0$-component of the CSM class, see (2.2) of \cite{Ohmoto}.
Moreover, one can view the orbifold CSM class as in (2.2) and (2.4) of \cite{Ohmoto}, as a sum
$$ C^{orb}_*(X/G)=C_*(\pi_* {\bf 1}^{\Z^2}_{X, G}) =\pi_* \iota^* C^G_*({\bf 1}^{\Z^2}_{X, G}) =\sum_g \iota_g^* C_*(X^g/C(g)), $$
where $\pi$ maps $G$-invariant constructible functions on $X$ to constructible functions on $X/G$ and
$\iota^*$ is the homomorphism from $H^G_*(X)$ to $H_*(X)$ (inclusion of $X$ as fiber of $X\times_G EG \to BG$);
the last sum is over conjugacy classes of elements $g\in G$, and $C(g)$ is the centralizer of $g$. 
In the latter form, these classes can be
viewed (after reinterpreting them {\em co}homologically) as residing in the {\em delocalized equivariant cohomology}
$H^*(X,G)= \oplus H^*(X^g)^{C(g)}$, see \cite{BBM}. Delocalized equivariant cohomology for symmetric products
was considered in \cite{Zhou}. In the more general case of orbifolds $X^n/G_n$ with a wreath product
$G_n=G^n\rtimes S_n$, the delocalized equivariant cohomology is obtained as image under 
the Connes--Chern character of the $K$-theoretic construction of \cite{Wang}
recalled in \S \ref{FockSec} below. Delocalized equivariant cohomology is also the natural cohomology
for string theory on orbifolds, in the sense of \cite{DHVW}, \cite{VaWi}.

\medskip

Although the Chern-Schwartz-MacPherson classes are defined as homological Chern classes
of {\em singular} varieties, they still admit a Chern--Weil type formulation in terms of curvature
forms (currents), as shown in \cite{Fu}. The construction of Chern--Weil representatives is based
on an embedding of the singular variety $X$ in a smooth ambient variety $M$, and in universal
differential forms (currents) $\gamma_k$ obtained from the pullbacks to $\P(T^*M)$ of the
Chern classes $C_k(M)$ and the powers $\zeta^r$ of a $2$-form $\zeta$ on $\P(T^*M)$
determined by the property that, on the total space $\bS(T^*M)$ of the Hopf bundle
$\pi_H: \bS(T^*M) \to \P(T^*M)$ with fiber $S^1$, one has $\pi_H^*\zeta =d\beta$, with
$\beta$ the generator of the cohomology of the fiber $S^1$. More precisely, one has (\S 2 of \cite{Fu})
$$ (\sum_{r\geq 0} \zeta^r ) \wedge \pi^* C_*(T^*M) =\sum_k (-1)^{\dim M -k} \gamma_k, $$
which, using the Chern-Weil curvature forms for $C_*(T^*M)$ leads to Chern-Weil representatives 
for the $\gamma_k$ The Chern-Schwartz-MacPherson class of $X \subset M$ is then obtained as the current 
$C_*(X)$ whose pairing with a form $\omega$ is given by
$$ \langle C_k(X), \omega \rangle 
=\langle \P(N^*(X)), \gamma_k \wedge \omega \rangle =\int_{\P(N^*(X))}
\gamma_k \wedge \omega, $$
where $N^*(X)$ is the Legendrian conormal cycle defined in \S 1 of \cite{Fu} 
and $\P(N^*(X))=\pi_{\cH, \sharp}(N^*(X)\lfloor \beta)$, that is, $N^*(X)$ is 
the extension of $\P(N^*(X))$ to the total space of the Hopf bundle $\cH$,
$N^*(X)=\P(N^*(X)) \times_{\cH} [S^1]$. In case of a smooth variety $X$
one has $N^*(X)=(-1)^{\dim M - \dim X} [\P(\nu^*(X))]$ where $\nu^*(X)$ is the
conormal bundle.

\medskip
\subsection{The Fock space of orbifold symmetric products}\label{FockSec}

Following results of Segal for the equivariant $K$-theory of symmetric products
\cite{Segal}, Weiqiang Wang established in \cite{Wang} analogous results for the more general case of
wreath products $G\sim S_n= G^n\rtimes S_n$, for a finite group $G$ acting on 
a locally compact Hausdorff paracompact $G$-space $X$. 

We use the notation $G_n=G^n\rtimes S_n$ as in \cite{Wang}.
Let $K^\bullet_{G_n}(X^n)$ be the equivariant $K$-theory and let
$K^\bullet_{G_n,\C}(X^n)=K^\bullet_{G_n}(X^n)\otimes_\Z \C$.
The Fock space $\cF_G(X)$ is given by
\begin{equation}\label{FockGX}
\cF_G(X):= \oplus_{n\geq 0} q^n \, K^\bullet_{G^n\rtimes S_n,\C}(X^n),
\end{equation}
with $q$ a formal variable (which keeps count of the graded structure) and
with the term $n=0$ equal to $\C$.
It is proved in \cite{Wang} that the Fock space $\cF_G(X)$ has the following
properties:
\begin{itemize}
\item $\cF_G(X)$ is a graded connected Hopf algebra with multiplication defined by
\begin{equation}\label{Fockprod}
K^\bullet_{G_n,\C}(X^n) \otimes K^\bullet_{G_m,\C}(X^m) \stackrel{K}{\to} 
K^\bullet_{G_n \times G_m,\C}(X^{n+m}) \stackrel{Ind}{\to} 
K^\bullet_{G_{n+m},\C}(X^n),
\end{equation}
where the first map $K$ is the K\"unneth isomorphism and the second map $Ind$ is the
induction maps for subgroups; the comultiplication is given by 
\begin{equation}\label{Fockcoprod}
K^\bullet_{G_n,\C}(X^n) \stackrel{R}{\to} \oplus_{m=0}^n 
K^\bullet_{G_m \times G_{n-m},\C}(X^n) \stackrel{K^{-1}}{\to} 
\oplus_{m=0}^n  K^\bullet_{G_m,\C}(X^m) \otimes K^\bullet_{G_{n-m},\C}(X^{n-m}),
\end{equation}
where the first map $R$ is restriction from $G_n$ to subgroups $G_m \times G_{n-m}$
and the second map $K^{-1}$ is the inverse of the K\"unneth isomorphism.
\item As a graded algebra (graded over $\Z^+\times \Z/2\Z$), $\cF_G(X)$ is isomorphic to the algebra
\begin{equation}\label{SusyAlgK}
\cS( \oplus_{n\geq 1} q^n\, K^0_{G,\C}(X) ) \otimes \Lambda (\oplus_{n\geq 1} q^n\, K^1_{G,\C}(X)), 
\end{equation}
where $\cS$ denotes the symmetric algebra and $\Lambda$ the exterior algebra. The graded dimension
satisfies
\begin{equation}\label{dimqFock}
\dim_q \cF_G(X) = \sum_{n\geq 0} q^n \dim K_{G_n,\C}(X^n) = \frac{\prod_{\ell\geq 1} (1+q^\ell)^{\dim K^1_{G,\C}(X)}}
{\prod_{\ell\geq 1} (1- q^\ell)^{\dim K^0_{G,\C}(X)}}.
\end{equation}
\item The orbifold Euler characteristics (in the string theory sense) 
of the symmetric products have a generating function
\begin{equation}\label{chiorbGenFun}
\sum_{n\geq 1} \chi^{orb}(X^n,G_n)\, q^n = \prod_{\ell=1}^\infty (1-q^\ell)^{-\chi^{orb}(X,G)}.
\end{equation}
\item $\cF_G(X)$ is a free $\lambda$-ring generated by $K^\bullet_{G,\C}(X)$.
\end{itemize}

\smallskip

\begin{rem}\label{FockRem}{\rm 
This construction, applied to $X=\Sigma_{g'}$ and $G=\Gamma(g,\underline{\nu})/\Gamma_{g'}$
provides a Fock space $\cF_G(\Sigma_{g'})$ for our setting, with all the properties listed above, and
\begin{equation}\label{chiorbGenSymn}
\sum_{n\geq 1} \chi^{orb}(\Sigma_{g'}^n, G^n\rtimes S_n)\, q^n = 
\prod_{\ell=1}^\infty (1-q^\ell)^{-\chi^{orb}(\Sigma_{g'}, G)}.
\end{equation}
}
\end{rem}

\section{Higher twisted index theory}\label{IndexSec}

Let $\cE$ be an orbifold vector bundle on the good $2$-dimensional orbifold $\Sigma$. 
It defines a class $[\cE]$ in $K^\bullet_{orb}(\Sigma)$.
On the $n$-fold product $\Sigma^n$ we consider the orbifold bundle $\cE^{\boxtimes n}$.
This determines an orbifold vector bundle $\cE_n$ on the symmetric product $\Sym^n(\Sigma)$.
Let $\tilde\cE$ be the pull back of $\cE$ to $\H$ and $\tilde\cE^{\boxtimes n}$ the
corresponding bundle on $\H^n$. Similarly, we consider the pullback $\cE'$ to $\Sigma_{g'}$
and the bundle $\cE'^{\,\boxtimes n}$ on $\Sigma_{g'}^n$.
The class $[\cE_n]\in K^\bullet_{orb}(\Sym^n(\Sigma))$ corresponds to the classes 
$[\tilde\cE^{\boxtimes n}]$ and $[\cE'^{\,\boxtimes n}]$, respectively, under the identifications 
$K^\bullet_{orb}(\Sym^n(\Sigma))=K^\bullet_{\Gamma_n}(\H^n)=K^\bullet_{G_n}(\Sigma_{g'}^n)$.
Let $\dirac_{\tilde\cE}^+$ be the twisted Dirac operator on $\H$ and let $\nabla = d-i\eta$
be the hermitian connection on $\H$ with curvature $\nabla^2=i\omega$, where $\omega$ is
the $\Gamma$ invariant $2$-form defined by the magnetic field. The operator $\dirac_{\tilde\cE}^+ \otimes \nabla$
commutes with the projective action $(\Gamma,\sigma)$. Similarly, we write $\dirac_{\tilde\cE^{\boxtimes n}}^+$
for the twisted Dirac operator on $\H^n$ and we consider $\dirac_{\tilde\cE^{\boxtimes n}}^+ \otimes \nabla_n$
where $\nabla_n = d-i\zeta$ where $\zeta =\sum_j \eta_j$ is the 1-form of Lemma \ref{2nformsigma}. By
the argument of Lemma \ref{2nformsigma}, we see that $\dirac_{\tilde\cE^{\boxtimes n}}^+ \otimes \nabla_n$
commutes with the projective action $(\Gamma_n, \sigma_n)$. The analytic index is the image
under the twisted Kasparov assembly map $\mu_{\sigma_n}: K^\bullet_{orb}(\Sym^n(\Sigma))\to K_\bullet(C^*_r(\Gamma_n,\sigma_n))$ of \eqref{twistKasp},
\begin{equation}\label{anind}
{\rm Ind}_{(\Gamma_n,\sigma_n)}(\dirac_{\tilde\cE^{\boxtimes n}}^+ \otimes \nabla_n) =\mu_{\sigma_n}([\cE_n]),
\end{equation}
with the property that (\S 2.3 of \cite{MaMa1})
\begin{equation}\label{L2ind}
 \Ind_{L^2}(\dirac_{\tilde\cE^{\boxtimes n}}^+ \otimes \nabla_n)= 
\tr ({\rm Ind}_{(\Gamma_n,\sigma_n)}(\dirac_{\tilde\cE^{\boxtimes n}}^+ \otimes \nabla_n). 
\end{equation}

\smallskip

A cyclic $2$-cocycle on an algebra $\cR$ is a multilinear map $t: \cR \times \cR\times \cR \to \C$ satisfying
$$ t(a,b,c)=t(c,a,b)=t(b,c,a) $$
$$ t(ab, c, d) - t(a, bc, d) + t(a,b,cd) -t(da,b,c) =0. $$
A dense involutive subalgebra $\cR(\Gamma,\sigma) \subset C^*_r(\Gamma,\sigma)$, which contains the
twisted group ring $\C[\Gamma,\sigma]$ and is closed under holomorphic functional calculus is constructed
in \S 4 of \cite{MaMa3}, as the intersection of the domains of the powers $\delta^k$ of the derivation
$\delta=[D,\cdot]$ associated to the operator $D\, \delta_\gamma = \ell(\gamma) \delta_\gamma$
that multiplies group elements by the word length $\ell(\gamma)$. The Haagerup inequality for
surface groups shows that group cocycles on $\Gamma$ with polynomial growth define
cyclic cocycles on $\C[\Gamma,\sigma]$ that extend continuously to $\cR(\Gamma,\sigma)$, see \cite{MaMa3}.
In particular, given a bounded $2$-cocycle $c\in Z^2(\Gamma)$, one has an associated cyclic $2$-cocycle $\tr_c$
on $\cR(\Gamma,\sigma)$. This gives an additive map on $K_0$, which we still denote by $\tr_c$.
Arguing as we did in \S \ref{magfieldSec} for the cocycle defined by the magnetic field, we can identify
$H^2(\Gamma)^{\oplus n}$ with a subspace of $H^2(\Gamma_n)$, as in Proposition \ref{sigmasigman}.
Let $c_n$ denote the cocycle in $H^2(\Gamma)^{\oplus n}$ defined by $n$ copies of the
$2$-cocycle $c \in H^2(\Gamma)$, and let $\tr_{c_n}$ be the corresponding cyclic $2$-cocycle
on $\cR(\Gamma_n,\sigma_n)$. 
As in \cite{CoMo} and in \S 3.2 of \cite{MaMa2}, one has an associated higher twisted analytic index
\begin{equation}\label{highanind}
{\rm Ind}_{(c_n,\Gamma_n,\sigma_n)}(\dirac_{\tilde\cE^{\boxtimes n}}^+ \otimes \nabla_n) =
\tr_{c_n} {\rm Ind}_{(\Gamma_n,\sigma_n)}(\dirac_{\tilde\cE^{\boxtimes n}}^+ \otimes \nabla_n) 
= \langle [\tr_{c_n}], \mu_{\sigma_n}([\cE_n])\rangle = \langle [c_n],[\cE_n]\rangle.
\end{equation}

\smallskip

We have the following index theorems (Theorem 1.1 of \cite{MaMa1} and Theorem 2.2 of \cite{MaMa2}), 
based on the Kawasaki index theorem on orbifold, \cite{Kawa1},
\cite{Kawa2}, see also \cite{Farsi}, and on the higher index theorem of \cite{CoMo}.

\begin{prop}\label{indextheorems}
Suppose given a cocycle $c\in H^2(\Gamma)$ as above, with $c_n$ the corresponding
cocyle on $\Gamma_n$, together with a multiplier $\sigma:\Gamma \times \Gamma \to U(1)$ as
in  \S \ref{magfieldSec}, determined by the closed $\Gamma$-invariant $2$-form $\omega$ of the magnetic field, and the corresponding $2$-form $\omega_n$ on $\Sigma_{g'}^n$. Then
the higher twisted index \eqref{highanind} is given by
\begin{equation}\label{kawaind}
\tr_{c_n} {\rm Ind}_{(\Gamma_n,\sigma_n)}(\dirac_{\tilde\cE^{\boxtimes n}}^+ \otimes \nabla_n) 
=\frac{1}{(2\pi)^n \, n! \, (\# G)^n}
\int_{\Sigma_{g'}^n} \hat A(\Omega_n) \tr(e^{R^{ \cE'^{\,\boxtimes n} }}) e^{\omega_n} \xi_{c_n},
\end{equation}
where $\xi_{c_n}$ is a $2$-form representative of the class in $H^2(\Sigma_{g'})^{\oplus n}$ corresponding
to $c_n \in H^2(\Gamma)^{\oplus n}\subset H^2(\Gamma_n)$, and $n! (\# G)^n=\# G_n$.
\end{prop}

The case \eqref{L2ind}, without the cyclic cocycle $c$, computes the range of the trace on $K$-theory,
which is useful for gap labelling purposes, see \cite{MaMa1}, \cite{MaMa3}. Here we focus on the
higher version with the cyclic cocycle, as that will provide the quantization of the Hall conductance
as in \cite{MaMa3}.

\smallskip

\begin{lem}\label{prodIndlem}
Let $\cE$ be an orbifold vector bundle over the good $2$-dimensional orbifold 
$\Sigma=\Sigma_{g'}/G=\H/\Gamma$ and $\cE'$ the pull back to $\Sigma_{g'}$. Given a
cocycle $c\in H^2(\Gamma)$ with the induced $c_n$ on $\Gamma_n=\Gamma^n\rtimes S_n$,
and let $\omega$ be the $2$-form determined by the magnetic field. Then the twisted higher
index theorem \eqref{kawaind} can be written as
\begin{equation}\label{prodInd}
\tr_{c_n} {\rm Ind}_{(\Gamma_n,\sigma_n)}(\dirac_{\tilde\cE^{\boxtimes n}}^+ \otimes \nabla_n) 
=\frac{1}{(2\pi)^n \, n! \, (\# G)^n} \left(
\int_{\Sigma_{g'}} \hat A(\Omega) \tr(e^{R^{ \cE'}}) e^\omega \xi_c \right)^n.
\end{equation}
\end{lem}

\proof The $\hat A$-genus is multiplicative over products, and the form $\hat A(\Omega_n)$ on
$\Sigma_{g'}^n$ is the product of $n$ copies of $\hat A(\Omega)$ on $\Sigma_{g'}$. The
Chern character ${\rm ch}(\cE)=\tr(e^{R^{ \cE}})$ is multiplicative for external tensor products,
hence $\tr(e^{R^{ \cE'^{\,\boxtimes n} }})$ on $\Sigma_{g'}^n$ is also a product of $n$ copies
of  $\tr(e^{R^{ \cE'}})$ each depending only on one of the factors $\Sigma_{g'}$. The form
$\omega_n$ is by construction (see \S \ref{magfieldSec}) a sum of $n$ copies of the $2$-form
$\omega$, each depending only on the coordinates of one of the $\Sigma_{g'}$ factors, hence
$e^{\omega_n}$ is also a product. Moreover, as we have seen above,
the $2$-form $\xi_{c_n}$ on $\Sigma_{g'}^n$ is also a product of copies of a $2$-form $\xi_c$
on $\Sigma_{g'}$. Thus, the integrand in \eqref{kawaind} splits as a product of identical
terms depending on only one of the factors.
\endproof

\smallskip

The area $2$-cocycle $c\in H^2(\Gamma)$
is the restriction to $\Gamma \subset \PSL(2,\R)$ of the hyperbolic area $c:  \PSL(2,\R)\times  \PSL(2,\R)\to \R$, where $c(\gamma_1,\gamma_2)$ is the oriented hyperbolic area of the geodesic
triangle in $\H$ with vertices $(z_0, \gamma_1^{-1} z_0, \gamma_2 z_0)$, for a chosen
base point $z_0$.

\smallskip

\begin{cor}\label{valuesInd}
In the case where $c\in H^2(\Gamma)$ is the area $2$-cocycle,
the range of values of the twisted higher index theorem, while varying the
choice of the orbifold vector bundle $\cE$ on $\Sigma$ is given by
\begin{equation}\label{rangeInd}
\chi^{orb}(\Sym^n(\Sigma))\,  \Z \subset \Q ,
\end{equation}
where $\chi^{orb}(\Sym^n(\Sigma))$ is the Satake orbifold Euler characteristic of $\Sym^n(\Sigma)$.
\end{cor}

\proof The integral
$$ \frac{1}{2\pi \# G} \int_{\Sigma_{g'}} \hat A(\Omega) \tr(e^{R^{ \cE'}}) e^\omega \xi_c $$
is the twisted higher index theorem computed in
\cite{MaMa2}. It is shown in Corollary 3.2 of \cite{MaMa2} that, when 
$c\in H^2(\Gamma)$ is the area $2$-cocycle, this integral is given by
$$ \frac{\chi(\Sigma_{g'})}{\# G} \cdot {\rm rank}(\cE) =\chi^{orb}(\Sigma)\cdot {\rm rank}(\cE) . $$
Thus, for the area cocycle, the twisted higher index theorem \eqref{prodInd} is given by
$$ \frac{\chi(\Sigma_{g'})^n}{n! (\#G)^n} \cdot {\rm rank}(\cE)^n =
\chi^{orb}(\Sym^n(\Sigma)) \cdot {\rm rank}(\cE)^n. $$
\endproof

\smallskip

The Hall conductance on $\Sigma$ is also described by a cyclic $2$-cocycle on the
twisted group algebra $\C[\Gamma,\sigma]$, given by
\begin{equation}\label{Hallcocycle}
\tr_K(f_0, f_1,f_2)=\sum_{j=1}^g \tr(f_0 (\delta_j(f_1)\delta_{j+g}(f_2)-\delta_{j+g}(f_1)\delta_j(f_2))),
\end{equation}
where $g$ is the genus and the $\delta_j$, for $j=1,\ldots,2g$ are derivations associated to
the elements of a symplectic basis of $H^1(\Sigma,\R)$. If $P_E$ denotes the spectral projection
associated to the Fermi level, then the Hall conductance is given by
$$ \sigma_E = \tr_K(P_E,P_E,P_E). $$
A derivation of this expression for the Hall conductance can be obtained as a quantum
adiabatic limit, see \cite{CHMM}.
As shown in Theorem 4.1 of \cite{MaMa2}, the conductance cocycle and the area cocycle
are cohomologous. Since the twisted higher index theorem, seen as a pairing of cyclic
cohomology and K-theory, only depends on the class of the cyclic cocycle, the range of
the twisted higher index theorem also determines the possible range of values of the
Hall conductance. We summarize the conclusion of this section as follows:
the single particle theory on $\Sigma$ with the external magnetic field $\omega$ extends
in a compatible way to a many particles model on the symmetric products $\Sym^n(\Sigma)$.
In this model, the range of quantized values of the Hall conductance consists of integer multiples
of the Satake orbifold Euler characteristics $\chi^{orb}(\Sym^n(\Sigma))$.

\medskip
\section{Orbifold braid groups and anyons}\label{AnyonsSec}

In this section we analyze what types of anyons and composite fermions 
one obtains within this model of fractional quantum Hall effect. These
are related to a notion of ``orbifold braid groups" that we introduce below.

\smallskip

The configuration space of $n$ (ordered) points on $\H$ is given
by the complement of the diagonals $F(\H,n)=\H^n\smallsetminus \Delta$.
The configuration spaces of unordered points is defined as the quotient
by the action of the symmetric group
\begin{equation}\label{FHn}
{\rm Conf}(\H,n):= (\H^n\smallsetminus \Delta)/S_n =F(\H,n)/S_n .
\end{equation}
These have fundamental group $\pi_1(\Conf(\H,n))=B_n$, the Artin braid group,
with generators $\sigma_i$, $i=1,\ldots, n-1$ and relations $\sigma_i\sigma_j=\sigma_j\sigma_i$
for $|i-j|\geq 2$ and $\sigma_i \sigma_{i+1} \sigma_i= \sigma_{i+1} \sigma_i \sigma_{i+1}$,
for $i=1,\ldots, n-2$. In fact, the spaces $\Conf(\H,n)$ are topologically Eilenberg-MacLane spaces
$K(B_n,1)$, see \cite{FaNeu}.

Given a $2$-dimensional compact (topological) surface $\Sigma$, and a finite set of points
$Q=\{ x_j \}_{j=1\ldots, m}$ on $\Sigma$, one similarly defines the configuration spaces
$$ F(\Sigma\smallsetminus Q, n) =(\Sigma\smallsetminus Q)^n \smallsetminus \Delta $$
\begin{equation}\label{ConfSigma}
{\rm Conf}(\Sigma\smallsetminus Q, n)=F(\Sigma\smallsetminus Q, n)/S_n .
\end{equation}
For $r<n$, the projections $\Pi_{n,r}(z_1,\ldots,z_n)=(z_1,\ldots,z_r)$ define locally trivial
fibrations $F(\Sigma\smallsetminus Q, n)\to F(\Sigma\smallsetminus Q, r)$ with the
fiber over $w=(w_1,\ldots,w_r)$ given by the configuration space
$F(\Sigma\smallsetminus (Q\cup\{w_i\}_{i=1,\ldots,r}), n-r)$. The braid group of $\Sigma\smallsetminus Q$
on $n$ strings is given by the fundamental group
\begin{equation}\label{braidSigma}
B_n(\Sigma\smallsetminus Q) := \pi_1({\rm Conf}(\Sigma\smallsetminus Q, n)).
\end{equation}
In particular, if $\Sigma$ is a $2$-dimensional orbifold and $Q\subset \Sigma$ is the
set of cone points, we have corresponding braid groups
\begin{equation}\label{braidorbSigma}
B_n(\Sigma_{\rm reg}) = \pi_1({\rm Conf}(\Sigma_{\rm reg}, n)) =\pi_1 ((\Sigma^n_{\rm reg}\smallsetminus \Delta)/S_n).
\end{equation}

\medskip
\subsection{Orbifold braid group}

For a good $2$-dimensional orbifold $\Sigma$, we can also associate to the configuration space 
${\rm Conf}(\Sigma_{\rm reg}, n)$ an {\em orbifold braid group}, defined as
\begin{equation}\label{orbBn}
B_n^{orb}(\Sigma) := \pi_1^{orb}({\rm Conf}(\Sigma, n)),
\end{equation}
using the orbifold fundamental group of ${\rm Conf}(\Sigma, n)$.

\begin{prop}\label{orbBnH}
The orbifold braid group of $\Sigma$ is a quotient of the ordinary braid group of $\Sigma_{\rm reg}$
by a normal subgroup generated by powers $\gamma_j^{\nu_j}$, with $\nu_j$ the order of the
stabilizer of the $j$-th cone point $x_j$ of $\Sigma$, and with $\gamma_j$ a loop
in ${\rm Conf}(\Sigma_{\rm reg}, n)$ that winds around the $j$-th component of the
(real) codimension two stratum of ${\rm Conf}(\Sigma, n)_{\rm sing}$.
\end{prop}

\proof
We have ${\rm Conf}(\Sigma_{\rm reg}, n)=(\Sigma^n_{\rm reg}\smallsetminus \Delta)/S_n \subset \Sym^n(\Sigma)_{\rm reg}$, as in Lemma \ref{pi1orbSn}, where
$$ (\Sigma^n\smallsetminus \Delta)_{\rm reg}=(\Sigma^n\smallsetminus \Delta)\cap \Sigma^n_{\rm reg}=\{ (z_1,\ldots,z_n)\in 
\Sigma^n_{\rm reg}\,|\, z_i\neq z_j \, \, \forall i, \} $$ $$ =\{ (z_1,\ldots,z_n)\in 
\Sigma^n\,|\, z_i\in \Sigma_{\rm reg}\, \, \text{ and } x_i\neq x_j \, \, \forall i,j\} =
\Sigma^n_{\rm reg}\smallsetminus \Delta. $$
Moreover, $((\Sigma^n_{\rm reg}\smallsetminus \Delta)/S_n)_{\rm reg}
=(\Sigma^n_{\rm reg}\smallsetminus \Delta)/S_n$, hence we can write the orbifold fundamental group as
$$ \pi_1^{orb}({\rm Conf}(\Sigma, n))= \pi_1({\rm Conf}(\Sigma_{\rm reg}, n))/H, $$
where $H$ is the normal subgroup generated by elements $\gamma_a^{\nu_a}$, where the
$\gamma_a$ are loops in ${\rm Conf}(\Sigma_{\rm reg}, n)$ around a component $X_a$ of the
singular locus ${\rm Conf}(\Sigma, n)_{\rm sing}$ with $\nu_a$ the order of the stabilizer of $X_a$.
The singular locus 
${\rm Conf}(\Sigma, n)_{\rm sing}=(\Sigma^n\smallsetminus \Delta)_{\rm sing}/S_n$ only comes
from the cone points of $\Sigma$, namely
$$ (\Sigma^n\smallsetminus \Delta)_{\rm sing}=(\Sigma^n)_{\rm sing} \smallsetminus \Delta, $$
where $(\Sigma^n)_{\rm sing}=\cup_{k=1}^n \Sigma_{{\rm sing},k}$, with
$$ \Sigma_{{\rm sing},k}=\Sigma \times \cdots \times \Sigma \times \Sigma_{\rm sing}\times \Sigma\cdots
\times \Sigma, $$
with a copy of $\Sigma_{\rm sing}$ in the $k$-th factor and $\Sigma$ in all the other factors. 
We denote by $\Sigma_{{\rm sing},k}(x_j) \subset \Sigma_{{\rm sing},k}$ the component 
of the (real) codimension two stratum of $(\Sigma^n)_{\rm sing}$
that has a cone point $\{ x_j \} \subset \Sigma_{\rm sing}$ in the $k$-th factor.
Thus, the components of the (real) codimension two stratum 
of $(\Sigma^n)_{\rm sing} \smallsetminus \Delta$ are of the form
$\Sigma_{{\rm sing},k}(x_j) \smallsetminus \Delta$. Let $\gamma_{j,k}$ be a loop in 
$\Sigma^n\smallsetminus \Delta$ that winds around the component $\Sigma_{{\rm sing},k}(x_j)$.
The power $\gamma_{j,k}^{\nu_j}$, where $\nu_j$ is the order of the stabilizer of the cone
point $x_j$ in $\Sigma$ is a generator of the subgroup $H$ of $\pi_1(\Sigma^n\smallsetminus \Delta)$
such that 
$$ \pi_1^{orb}(\Sigma^n\smallsetminus \Delta)=\pi_1(\Sigma^n\smallsetminus\Delta)/H. $$
Passing to the quotient $F(\Sigma,n)=\Sigma^n\smallsetminus \Delta \to \Conf(\Sigma,n)
=(\Sigma^n\smallsetminus \Delta)/S_n$,  the (real) codimension two stratum of the singular
locus is the image of the $S_n$ invariant configuration of components in $F(\Sigma,n)$
given by $\Sigma^n_{\rm sing}(x_j):=\cup_{k=1}^n \Sigma_{{\rm sing},k}(x_j)$, 
for a given cone point $x_j$. We denote
by $\gamma_j$ a loop in $\Conf(\Sigma,n)$ winding around $\Sigma^n_{\rm sing}(x_j)$.
The powers $\gamma_j^{\nu_j}$ generate the subgroup $H$ with
$$ \pi_1^{orb}(\Conf(\Sigma,n))=\pi_1(\Conf(\Sigma_{\rm reg},n))/H. $$
\endproof

\smallskip

The braid groups of an $2$-dimensional orientable (topological) surface of genus $g$ 
with $m$ punctures were computed explicitly in \cite{Birman1}, \cite{Birman2} (see
also \cite{Scott2}, \cite{Belling} for a slightly different form of the presentation). With
the presentation given in \cite{Belling}, if $\Sigma$ has genus $g$ with a set 
$Q=\{ x_j \}_{j=1,\ldots, m}$ of $m$ punctures, the braid group $B_n(\Sigma\smallsetminus Q)$
has additional generators with respect to the Artin braid group $B_n$. Namely, the gerators
are given by  
\begin{equation}\label{braidSigmaQgrp}
\begin{array}{ll}
\sigma_i, &  i=1,\ldots n-1 \\
a_\ell , & \ell=1,\ldots,g \\
b_\ell, & \ell=1,\ldots,g \\
c_j, & j=1,\ldots, m-1
\end{array}
\end{equation}
with relations
\begin{equation}\label{braidrels}
\begin{array}{ll}
\sigma_i\sigma_j=\sigma_j\sigma_i, & \text{ when } |i-j|\geq 2 \\
\sigma_i \sigma_{i+1} \sigma_i= \sigma_{i+1} \sigma_i \sigma_{i+1}, & i=1,\ldots, n-2,
\end{array}
\end{equation}
and in the Artin braid group, and additional relations, for all $\ell$, 
\begin{equation}\label{braidrel2}
\begin{array}{ll}
a_\ell \sigma_i =\sigma_i a_\ell, &  i\neq 1, \\
b_\ell \sigma_i =\sigma_i b_\ell, &  i\neq 1,
\end{array}
\end{equation}
\begin{equation}\label{braidrel3}
\begin{array}{cl}
\sigma_1^{-1} a_\ell \sigma_1^{-1} a_\ell= a_\ell \sigma_1^{-1} a_\ell \sigma_1^{-1} &  \\
\sigma_1^{-1} b_\ell \sigma_1^{-1} b_\ell= b_\ell \sigma_1^{-1} b_\ell \sigma_1^{-1} &  \\
\sigma_1^{-1} a_\ell \sigma_1^{-1} b_\ell= b_\ell \sigma_1^{-1} a_\ell \sigma_1 &  \\
\sigma_1^{-1} a_\ell \sigma_1 a_r= a_r \sigma_1^{-1} a_\ell \sigma_1 &  \ell< r, \\
\sigma_1^{-1} b_\ell \sigma_1 b_r= b_r \sigma_1^{-1} b_\ell \sigma_1 & \ell< r , \\
\sigma_1^{-1} a_\ell \sigma_1 b_r= b_r \sigma_1^{-1} a_\ell \sigma_1 & \ell< r , \\
\sigma_1^{-1} b_\ell \sigma_1 a_r= a_r \sigma_1^{-1} b_\ell \sigma_1 &  \ell< r , \\
\end{array}
\end{equation}
and for all $\ell,r$ and all $j$
\begin{equation}\label{braidrel4}
\begin{array}{cl}
c_j \sigma_i = \sigma_i c_j, & i\neq 1,  \\
 \sigma_1^{-1} c_j \sigma_1 a_r= a_r \sigma_1^{-1} c_j \sigma_1 & n>1,\\
\sigma_1^{-1} c_j \sigma_1 b_r= b_r \sigma_1^{-1} c_j \sigma_1 & n>1, \\
\sigma_1^{-1} c_j \sigma_1 c_k= c_k \sigma_1^{-1} c_j \sigma_1 &  j<k, \\
\sigma_1^{-1} c_j \sigma_1^{-1} c_j= c_j \sigma_1^{-1} c_j \sigma_1^{-1} .&  \\ 
\end{array}
\end{equation}
In the case without punctures, the braid group $B_n(\Sigma)$ has generators
$\sigma_i$, $a_\ell$, $b_\ell$ as above with the same relations \eqref{braidrels},
\eqref{braidrel2}, \eqref{braidrel3}, but with \eqref{braidrel4} replaced by 
\begin{equation}\label{braidrel5}
\prod_{\ell=1}^g [a_\ell, b_\ell^{-1}] = \sigma_1 \sigma_2 \cdots \sigma_{n-1}^2 \cdots \sigma_2 \sigma_1.
\end{equation}

\smallskip

\begin{prop}\label{orbBnpres}
Let $\Sigma$ be a $2$-dimensional good orbifold of genus $g$,
with $m$ cone points. The orbifold braid groups $B_n^{orb}(\Sigma)$
has generators $\sigma_i$, $i=1,\ldots, n-1$, $a_\ell$, $b_\ell$, $\ell=1,\ldots, g$
and $c_j$, $j=1,\ldots,m$ with relations as above, and with the additional relation 
\begin{equation}\label{orbbraidrel5}
 \prod_{\ell=1}^g [a_\ell, b_\ell^{-1}]\,\,  \sigma_1^{-1} \sigma_2^{-1} \cdots \sigma_{n-1}^{-2} \cdots \sigma_2^{-1} \sigma_1^{-1} \,\, c_1 \cdots c_m =1 
\end{equation}
and $c_j^{\nu_j}=1$, where $\nu_j$ is the order of the stabilizer of the cone point.
\end{prop}

\proof 
As shown in \cite{Belling}, the generators $a_\ell$ and $b_\ell$ correspond geometrically, in
terms of a fundamental domain for $\Sigma$ given by a $4g$-gons with pairwise identified
sides marked by the generators $\alpha_\ell$ and $\beta_\ell$ of $\pi_1(\Sigma)$. The braid $a_\ell$
is a string that crosses the $\alpha_\ell$ sides and $b_\ell$ the $\beta_\ell$ side (with the opposite
orientation), while the $c_j$ wind around the $j$-th puncture. The generators $\sigma_i$ have the 
usual meaning as in the Arting braid group.
The relations are explained geometrically in \S 2.2 of \cite{Belling}, where it is also shown
that one can equivalently introduce an additional generator $c_m$, as the braid that winds 
around the last puncture, and the additional relation \eqref{orbbraidrel5}. This corresponds
to writing the fundamental group of the punctured surface $\Sigma\smallsetminus Q=\Sigma_{\rm reg}$ as
$$ \pi_1(\Sigma\smallsetminus Q)=
\langle \{ a_\ell, b_\ell \}_{\ell=1,\ldots,g}, \{ c_j \}_{j=1,\ldots,m}\,
 \,|\, \prod_\ell [a_\ell, b_\ell^{-1}] c_1 \cdots c_m =1 \rangle, $$
instead of writing it solely in terms of the generators $a_\ell, b_\ell, c_j$ with $j=1,\ldots, m-1$.
In particular, the generators $c_j$ 
provide loops in the configuration space $\Conf(\Sigma_{\rm reg},n)$ that wind around 
the component $\Sigma^n_{\rm sing}(x_j)$. Thus, by Proposition \ref{orbBnH}
we obtain $B^{orb}_n(\Sigma)$ from $B_n(\Sigma_{\rm reg})$
by imposing the further relations $c_j^{\nu_j}=1$.
\endproof

\medskip
\subsection{Anyons}

Fractional-statistics particles, or anyons, have the property that, when two
particles get interchanged, the wavefunction changes by a phase factor
$\exp(i \pi \alpha)$, for some $\alpha \in (-1, 1]$. The cases $\alpha=0$
and $\alpha=1$ correspond, respectively, to bosons and fermions.
It is known \cite{Einar}, \cite{Imbo} (see also \cite{Lerda}) that, for 2-dimensional systems, the
surface topology plays an important role in determining what type of
anyons can arise. More precisely, these are classified by $1$-dimensional
unitary representations of the braid group of the surface.  The case of the
orbifold braid group is similar.

\begin{lem}\label{fermionsonly}
Let $\Sigma$ be a $2$-dimensional good orbifold of genus $g>0$,
with $m$ cone points. One-dimensional unitary representations $R$ of 
the orbifold braid group $B_n(\Sigma)$ have the generators $\sigma_i$
acting as $R(\sigma_i)=\pm 1$, the generators $a_\ell$ and $b_\ell$, respectively, acting
as phase factors $R(a_\ell)=e^{2\pi i \theta_\ell}$ and $R(b_\ell)=e^{2\pi i \phi_\ell}$
and the generators $c_j$ acting as $R(c_j)=e^{2\pi i \beta_j/\nu_j}$, where 
$\nu_j$ is the order of the stabilizer of the $j$-th cone point and the $\beta_j$
satisfy $\sum_{j=1}^m \beta_j/\nu_j \in \Z$.
\end{lem}

\proof The fact that the $\sigma_i$ must act like $\pm 1$ follows from the relation
$\sigma_1^{-1} a_\ell \sigma_1^{-1} b_\ell= b_\ell \sigma_1^{-1} a_\ell \sigma_1$
in \eqref{braidrel3}, which implies that $\sigma_1$ acts as $\pm 1$ and the
relations \eqref{braidrels}, which imply that all the $\sigma_i$ must then also
act in the same way. The action of the $a_\ell$ and $b_\ell$ is unconstrained
by the relations, hence we get independent phase factors for each of them,
while the $c_j$ are constrained by the relations $c_j^{\nu_j}=1$ and \eqref{orbbraidrel5},
which implies $R(c_1\cdots c_m)=1$. These give $R(c_j)=e^{2\pi i \beta_j/\nu_j}$
with $\sum_{j=1}^m \beta_j/\nu_j \in \Z$.
\endproof

\begin{rem}\label{anyonrem1}{\rm 
Since the braids $\sigma_i$ correspond to exchanging two particles,
all the representations in Lemma \ref{fermionsonly} are either fermions
or bosons, whenever $g>0$. 
The $\beta_j$ can be viewed as the Seifert invariants of an orbifold line bundle
over $\Sigma$ that has integer orbifold Euler number (hence is an actual
line bundle). See \S \ref{SeifertSec} for more details.}
\end{rem}

In the case of good $2$-dimensional orbifolds 
of genus $g=0$ with $m$ cone points, we have the following result.

\begin{lem}\label{g0orbanyons}
Let $\Sigma$ be a $2$-dimensional good orbifold of genus $g=0$,
with $m$ cone points. Then the one-dimensional unitary representations 
$R$ of the orbifold braid group $B_n(\Sigma)$ have $R(\sigma_i)=e^{i\pi\alpha}$
and $R(c_j)=e^{2\pi i \beta_j/\nu_j}$, where 
$\nu_j$ is the order of the stabilizer of the $j$-th cone point and the $\beta_j$
satisfy 
\begin{equation}\label{betajalpha}
\alpha - \sum_{j=1}^m \frac{\beta_j}{\nu_j} \,\,\,  \in \Z.
\end{equation}
\end{lem}

\proof The argument is exactly as in the previous lemma, except that
we do not have the generators $a_\ell$, $b_\ell$ and the only relations
are the \eqref{braidrels} and
$$ \begin{array}{cl}
c_j \sigma_i = \sigma_i c_j, & i\neq 1 \\
\sigma_1^{-1} c_j \sigma_1 c_k= c_k \sigma_1^{-1} c_j \sigma_1 &  j<k, \\
\sigma_1^{-1} c_j \sigma_1^{-1} c_j= c_j \sigma_1^{-1} c_j \sigma_1^{-1} &  \\  
\sigma_1 \sigma_2 \cdots \sigma_{n-1}^{2} \cdots \sigma_2 \sigma_1= c_1 \cdots c_m & 
\end{array}
$$
and $c_j^{\nu_j}=1$. This last relation gives, as before, $R(c_j)=e^{2\pi i \frac{\beta_j}{\nu_j}}$.
The relations \eqref{braidrels} imply that all the $\sigma_i$ must act by the same
phase factor $R(\sigma_i)=e^{i\pi \alpha}$ and the last displayed relation then implies
that $e^{2\pi i\alpha}=e^{2\pi i \sum_{j=1}^m \beta_j/\nu_j}$, hence we obtain \eqref{betajalpha}.
\endproof

\begin{rem}\label{anyonrem2}{\rm 
Thus, in the case of good orbifolds of genus zero, there are non-trivial anyons
(that are neither fermions nor bosons) and the fractional statistics they satisfy 
depends on the datum of an orbifold line bundle on $\Sigma$, through the 
Seifert invariants $\beta_j$, see \S \ref{SeifertSec} below.}
\end{rem}

\smallskip

Moreoever, it is known that two-dimensional systems on surfaces of genus $g>0$ 
do admit fractional statistics arising from higher dimensional irreducible 
unitary representations of the braid group $B_n(\Sigma)$, 
\cite{Einar}, \cite{Imbo}, provided $\alpha$ satisfies $\exp(2\pi i(n+g-1)\alpha)=1$.
We describe the analog for the orbifold braid group.

\smallskip

Consider the $N\times N$ matrices
\begin{equation}\label{UN}
U_N =\left(\begin{array}{ccccc} 1 & & & & \\
& \xi_N^2 & & & \\
& & \xi_N^4 & & \\
 & & & \ddots & \\
& & & & \xi_N^{2(N-1)}
\end{array}\right), \ \ \ \text{ with } \ \  \ \xi_n =\exp(\pi i /N),
\end{equation}
\begin{equation}\label{Vn}
V_N =\left(\begin{array}{cccccc} 
0 & 1 & 0 & \cdots & 0 & 0 \\
0 & 0 & 1 & \cdots & 0 & 0 \\
\vdots & & & & & \vdots \\
0 & 0 & 0 & \cdots & 0 & 1 \\
1 & 0 & 0 & \cdots & 0 & 0
\end{array}\right).
\end{equation}
They satisfy the commutation relation 
\begin{equation}\label{UVN}
V_N\, U_N = \xi_N^2\, \, U_N\, V_N.
\end{equation}

\smallskip

\begin{prop}\label{highdimanyons}
Let $\Sigma$ be a good $2$-dimensional orbifold of genus $g$ with $m$ cone points.
The orbifold braid group $B_n^{orb}(\Sigma)$ has  
unitary representations of dimension $N^g$ with $R(\sigma_i)=\xi_N^{-1}$ 
and $R(a_\ell)=U_{N,\ell}$ and $R(b_\ell)=V_{N,\ell}$,
where $U_{N,\ell}$ and $V_{N,\ell}$ act as $U_N$ and $V_N$, respectively, in the $\ell$-th
factor of the tensor product of $g$-copies of $\C^N$ and the identity on the other factors. 
The generators $c_j$ act as $R(c_j)=e^{2\pi i \beta_j/\nu_j}$, where $\nu_j$ is the 
order of the stabilizer of the $j$-th cone point and the $\beta_j$ satisfy the relation
\begin{equation}\label{betajNgdim}
\frac{(g+n-1)}{N} + \sum_{j=1}^m \frac{\beta_j}{\nu_j} \,\,\,  \in \Z.
\end{equation}
\end{prop}

\proof The relation 
$\sigma_1^{-1} a_\ell \sigma_1^{-1} b_\ell= b_\ell \sigma_1^{-1} a_\ell \sigma_1$
is now satisfied, since the left-hand-side $\xi_N^2 U_{N,\ell} V_{N,\ell}$
and the right-hand-side $V_N U_N$ agree by \eqref{UVN}. The rest of the
relations \eqref{braidrel3} are also satisfied, since for $\ell\neq r$
$U_{N,\ell}$ and $V_{N,r}$ commute. The relation $c_j^{\nu_j}=1$ is satisfied by
$R(c_j)=e^{2\pi i \beta_j/\nu_j}$, and the rest of the relations \eqref{braidrel4}
are also satified. The remaining relation \eqref{orbbraidrel5} for the orbifold
braid groups implies
$$ \prod_{\ell=1}^g R([a_\ell,b_\ell^{-1}]) \, \prod_i R(\sigma_i)^{-2} \, R(c_1\cdots c_m) =1 $$
Note that from \eqref{UVN} we have $U_{N,\ell} V_{N,\ell}^{-1} = \xi_N^2  V_{N,\ell}^{-1} U_{N,\ell}$.
Thus, we obtain
$$ \xi_N^{2g} \cdot \xi_N^{2(n-1)} \cdot e^{2\pi i \sum_{j=1}^m \frac{\beta_j}{\nu_j}} =1, $$
which gives $\exp(2\pi i ((g+n-1)/N + \sum_j \beta_j/\nu_j))=1$, namely \eqref{betajNgdim}.
\endproof

\medskip
\subsection{Orbifold line bundles and orbifold Euler numbers}\label{SeifertSec}

We clarify here the relation (mentioned in Remarks \ref{anyonrem1} and \ref{anyonrem2}) 
between the anyon representations described above and the Seifert invariants of
orbifold line bundles.
For a complex vector bundle $\cE$ of rank $n$ over a manifold $X$ of real dimension $2n$,
the Euler number $\chi(\cE)$ is the integral on $X$ of the Euler class $e(\cE)$. In the
case of a line bundle on a $2$-dimensional surface, the Euler number is the integral of
the first Chern class. 
For an orbifold line bundle $\cL$ on a (good) $2$-dimensional orbifold $\Sigma$, 
the Euler number $\chi(\cL)$ is replaced by an {\em orbifold
Euler number} (see \cite{Scott}, p.437)
\begin{equation}\label{orbchiE}
\chi^{orb}(\cL)= \chi(\cL) - \sum_{j=1}^m \frac{\beta_j}{\nu_j}, 
\end{equation}
where the Euler number $\chi(\cL)$ 
is corrected by a contribution for each cone point $x_j$, $j=1,\ldots, m$
of the orbifold. These corrections are of the form $\beta_j/\nu_j$, where $\nu_j$ is
the order of the stabilizer $\Z/\nu_j \Z$ of the cone point $x_j$ and the $\beta_j$ are 
the Seifert invariants of the orbifold line bundles. These are obtained by considering the
associated principal $U(1)$-bundle $P$ and the exact sequence
$$ 1 \to \Z \to \pi_1(P) \to \Gamma \to 1, $$
where if $c_j$ is one of the generators of $\Gamma$ with $c_j^{\nu_j}=1$ and $\alpha$
is the generator of the fundamental group $\Z$ of the fiber, then the $\beta_j$, with
$0\leq \beta_j \leq \nu_j-1$, are defined
by the relation $$\alpha^{\beta_j} =\tilde c_j,$$ where $\tilde c_j$ is a preimage of $c_j$ in
$\pi_1(P)$. By the Hopf theorem, the Euler number of a line bundle on a $2$-dimensional
surface is a sum over zeros of a section of the line bundle, counted with 
multiplicity. One can then think of the orbifold Euler number \eqref{orbchiE} as a modification of this counting,
where additional zeros are counted at the cone points, with multiplicities $\beta_j$,
but so that each zero only contributes a fraction $1/\nu_j$ of a zero at a regular point.
This is consistent with the Satake orbifold Euler characteristic $\chi^{orb}(\Sigma)$,
where vertices of a triangulation that are located at cone points are counted with a
factor of $1/\nu_j$. In fact $\chi^{orb}(\Sigma)$ is the orbifold Euler number of the
orbifold tangent bundle.

\medskip
\section{Laughlin-type wave functions}\label{LaughlinSec}

This section is more speculative in nature. It contains some observations on how one may
naturally encounter some Laughlin-type functions in the geometric setting
described in the previous sections.
The Laughlin wave function can be regarded as a generalization of the Slater function
\begin{equation}\label{Slater}
\Psi_{\rm Slater}(z_1,\ldots,z_n) = V(z_1,\ldots,z_n) \cdot e^{-\sum_{i=1}^n \frac{|z_i|^2}{4\ell^2}}, 
\end{equation}
with the Vandermonde determinant
$$ V(z_1,\ldots,z_n)= \prod_{1\leq i < j \leq n} (z_i - z_j) $$
and with the magnetic length $\ell =\sqrt{\frac{\hbar c}{eB}}$. The Slater function describes non-interacting
fermions in a magnetic field, for the full filling of the lowest Landau level. The Laughlin
wave function takes the form
\begin{equation}\label{Laughlin}
\Psi_{\rm Laughlin}(z_1,\ldots,z_n) = V(z_1,\ldots,z_n)^p \cdot e^{-\sum_{i=1}^n \frac{|z_i|^2}{4\ell^2}}, 
\end{equation}
so that it acquires a $p$-fold zero along the diagonals $z_i=z_j$. The exponent $p$ is
taken to be an odd integer, so that antisymmetry is preserved. In the case where $p$ is
an even integer, one considers functions of the form
\begin{equation}\label{Pfaffian}
\Psi_{\rm Pfaffian}(z_1,\ldots,z_n) = {\rm Pfaff}(\frac{1}{z_i-z_j})\cdot V(z_1,\ldots,z_n)^p \cdot e^{-\sum_{i=1}^n \frac{|z_i|^2}{4\ell^2}}.
\end{equation}
Explicit algorithmic methods for expressing even powers of the Vandermonde
determinant as combinations of Schur functions, and Laughlin wave functions
as combinations of Slater functions, are discussed in \cite{Balla}, \cite{SchThiWyb}.

\smallskip

By analogy with the expression $\prod_i (z_i -z)^p$ for a vortex of vorticity $p$ centered at $z$,
the Vandermonde determinant $V(z_1,\ldots, z_n)$ in the Slater wave function
can be thought of as describing cyclotron motion of $n$ non-interacting fermions
on the plane with magnetic field corresponding to a completely filled lower Laudau level $\nu=1$,
 and the corresponding powers $V(z_1,\ldots, z_n)^p$ in the Laughlin wave function
can then be thought of similarly as vortices with vorticity $p$, see \cite{JSWW}, \cite{JGJJ}.

\smallskip

We seek here some geometric interpretation of Laughlin type wave functions
related to the anyon representations described in the previous section and
the geometry of the orbifold symmetric products.

\smallskip
\subsection{Anyon representation and orbifold vector bundle}

We associate orbifold vector bundles to the anyon representations described in
the previous section.

\begin{lem}\label{VNanyon}
An $N$-dimensional anyon representations as in Proposition \ref{highdimanyons}
determines an orbifold line bundle $\cL$ on $\Sigma$, with pullback $\cL'$ to $\Sigma_{g'}$,
and an orbifold local system $V_N$ of (complex) rank $N$. These data in turn determine 
a rank $nN$ orbifold vector bundle over $\Conf(\Sigma,n)\subset \Sym^n(\Sigma)$
of the form $\cV_{n,N}=\cW_n \otimes V_N$, where $\cW_n$ is the restriction to
$\Conf(\Sigma,n)$ of the orbifold vector bundle on $\Sym^n(\Sigma)$ determined by
the external Whitney sum $\cL'^{\, \boxplus n}$ on $\Sigma_{g'}^n$.
\end{lem}

\proof By construction, the anyon representation is representation of $\pi_1^{orb}(\Conf(\Sigma,n))$
hence it determines an orbifold local system on $\Conf(\Sigma,n)$, in the same way as representations 
of the ordinary fundamental group define local systems. The Seifert data $\beta_j$ of the anyon
representation determine an orbifold line bundle $\cL$ over $\Sigma$. The product $\cW_n \otimes V_N$
is then an orbifold vector bundle over $\Conf(\Sigma,n)$ of complex rank $nN$. 
\endproof

The orbifold Chern number of $\cW_n$ is obtained as follows.

\begin{lem}\label{chiEorbG}
Let $\cL$ be an orbifold line bundle on the good $2$-dimensional
orbifold $\Sigma =\Sigma_{g'}/G$. Let $\cL'$ be the pullback $G$-equivariant line bundle on $\Sigma_{g'}$.
Consider the $n$-fold external Whitney sum $\cL'^{\, \boxplus n}$ over $\Sigma_{g'}^n$, and let
$\cW_n$ be the corresponding orbifold vector bundle over $\Sym^n(\Sigma)$.
Then the orbifold Euler number is given by
\begin{equation}\label{orbchiEboxn}
\chi^{orb}(\cW_n)= \frac{\chi(\cL')^n}{n! (\#G)^n} = \frac{1}{n!} \,\chi^{orb}(\cL)^n.
\end{equation}
\end{lem}

\proof
By Theorem 3.6 of \cite{Scott} the orbifold Euler number $\chi^{orb}(\cL)$ of an
orbifold line bundle on $\Sigma$ is related to the Euler number of a line bundle $\cL'$ 
on $\Sigma_{g'}$ that orbifold covers $\cL$ by
\begin{equation}\label{chiorbEcover}
\chi(\cL')=\chi^{orb}(\cL)\,\, \frac{\# G}{m},
\end{equation}
where $m$ is the number of times the circle in the fiber of the principal $U(1)$-bundle 
$P(\cL')$ wraps around the circle in the fiber of $P(\cL)$. When $m=1$, one obtains
$\chi^{orb}(\cL)=(\# G)^{-1} \chi(\cL')$. The external Whitney sum is the Whitney sum
$\oplus_{i=1}^n \pi_i^* \cL'$, where $\pi_i: \Sigma_{g'}^n \to \Sigma_{g'}$ is the
projection to the $i$-th factor. The Euler class of a Whitney sum is the cup product
of the Euler classes, hence we have $e(\cL'^{\, \boxplus n})=\bigwedge^n e(\cL')$
and the Euler number is $\chi(\cL'^{\, \boxplus n}) =\int_{\Sigma_{g'}^n} e(\cL'^{\, \boxplus n}) =
\chi(\cL')^n$. Finally, the relation between the orbifold Euler number of $\cW_n$ on the 
symmetric products $\Sym^n(\Sigma)$ and the Euler number of $\cL'^{\, \boxplus n}$ on $\Sigma_{g'}^n$ is
$$ \chi^{orb}(\cW_n) = \frac{1}{\# G_n} \, \chi(\cL'^{\,\boxtimes n}) =\frac{\chi(\cL')^n}{n! (\#G)^n} = \frac{1}{n!} \,\chi^{orb}(\cL)^n. $$
\endproof

Local systems have torsion Chern classes, hence they do not change the differential
form realizing the Euler class of $\cL'^{\,\boxtimes n}$ and its integration on $\Sigma_{g'}^n$.

\smallskip
\subsection{Mathai--Quillen formalism}

For a vector bundle $\cE$ of (real) rank $2n$ over a smooth manifold $X$ of (real) dimension $2n$,
the Euler class, whose integral $\chi(\cE)=\int_X e(\cE)$ computes the Euler number, 
is the pullback along the zero section of the bundle of a representative of the Thom class.
Pullbacks $e_s(\cE)$ along other sections give the same cohomology class. By Chern--Weil
theory, the Euler form can be written as the Pfaffian of the curvature $\Omega^\cE$ of a
hermitian connection on the bundle $\cE$,
$$ e(\cE) = \frac{1}{(2\pi)^m} {\rm Pfaffian}(\Omega^\cE). $$
The Pfaffian of an antisymmetric matrix can be written in terms of the Berezin integral
in fermionic coordinates
$$ {\rm Pfaffian}(A)= \int \cD \xi \, \exp(\frac{1}{2} \xi^i A_{ij} \xi^j), $$
hence one can write the Euler form as
$$ e(\cE) = \frac{1}{(2\pi)^m} \int \cD \xi \, \exp(\frac{1}{2} \xi^i \Omega^\cE_{ij} \xi^j). $$
An explicit representative for the Thom class, which is exponentially decaying along the
fibers (with normalizaed integral) and pulls back to the Euler form along the zero section
is given in \cite{MaQui} as
$$ \Phi_{MQ}(\cE)= \frac{-\eta^2/2}{(2\pi)^m} \int 
\cD \xi \, \exp(\frac{1}{2} \xi^i \Omega^\cE_{ij} \xi^j + i \nabla \eta^i\xi_i) , $$
where $\eta$ are the fiber coordinates. It is shown in \cite{MaQui} that this is indeed
a closed form representing the Thom class. The pullback along a nontrivial section gives
\begin{equation}\label{esE}
 e_s(\cE)= \frac{-s^2/2}{(2\pi)^m} \int 
\cD \xi \, \exp(\frac{1}{2} \xi^i \Omega^\cE_{ij} \xi^j + i \nabla s^i\xi_i) . 
\end{equation}
If the section $s$ is scaled by a factor $\lambda$, in the limit of large $\lambda \to \infty$
the form $e_s(\cE)$ localizes on the zero set $Z_s=\{ x\in X\,|\, s(x)=0 \}$ of the section,
hence recovering the Hopf theorem. The Mathai--Quillen formalism has found useful
applications in physics, based on the observation \cite{AtJe} that the partition function
of certain $N=1$ supersymmetric gauge theories can be written as formal functional integral
analog of $\int_X e_s(\cE)$.

\smallskip

\smallskip
\subsection{Vandermonde determinants and symmetric products}

For $S=(s_1,\ldots, s_n)$, let $e_j(S)$ be the $j$-th elementary symmetric function in these
variables, with 
$$ \prod_{j=1}^n (1+t s_j) = \sum_{j=0}^n t^j e_j(S). $$
Given $n$ symmetric polynomials $f_1, \ldots, f_n$, the Jacobian 
$$ J(f_1, \ldots, f_n)= \det\left( \frac{\partial f_i}{\partial s_j} \right), $$
while in the variables $e_j$, the Jacobian
$$ J_e (f_1, \ldots, f_n)= \det\left( \frac{\partial f_i}{\partial e_j} \right) $$
is related to $J(f_1, \ldots, f_n)$ by 
$$ J(f_1, \ldots, f_n)= J_e (f_1, \ldots, f_n) \cdot V, $$
where
$$ V(s_1,\ldots,s_n)= \prod_{1\leq i < j \leq n} (s_i - s_j) = J(e_1,\ldots, e_n) $$
is the Vandermonde determinant, see \cite{LasPra}, where explicit
expressions in terms of Schur functions
are given for the Jacobians $J_e$ of complete functions and
power sums. Thus, on the symmetric products $\Sym^n(\Sigma)$ we
should interpret the Vandermonde determinant $V(s_1,\ldots, s_n)$ as the
Jacobian of the change of local coordinates between the coordinates 
$(s_1,\ldots,s_n)$ of $\Sigma^n$ to the coordinates given by the symmetric 
functions $(e_1,\ldots, e_n)$. 

\smallskip
\subsection{Laughlin type functions from anyon representations}

Consider an $N$-dimensional anyon representation as in Proposition
\ref{highdimanyons} and the associated orbifold vector bundle $\cV_{n,N}$
on $\Conf(\Sigma,n)$, constructed as in Lemma \ref{VNanyon}. Let 
$\cL$ be the orbifold line bundle on $\Sigma$ determined
by the Seifert data of the anyon representation and $\cL'$ the pullback
to $\Sigma_{g'}$. Let $\underline{s}$ be a section of $\cL'^{\, \boxplus n}$ 
determined by a $n$-tuple of sections of $\cL'$, and let $e_{\underline{s}}(\cL'^{\, \boxplus n})$
be the Mathai--Quillen representative of the Euler class. For the orbifold vector bundle
$\cW_n$, integration in the fiber direction now takes place with respect to coordinates 
given by the elementary symmetric functions in the original coordinates. This can be
expressed in terms of the original coordinates by introducing the change of variables,
as above, in the form of the Vandermonde determinant $V(\underline{s})$. 
Thus, we obtain an expression that has a product of $V(\underline{s})$ combined with an
exponentially decaying factor in the fiber coordinates, as in the Slater wave function.
When we further tensor with the local system $V_N$, each block of $n$-coordinates in the
measure along the fiber directions acquires a factor equal to the Vandermonde determinant, 
while the Euler class is unchanged, hence producing a product of a power
$V(\underline{s})^N$ of the Vandermonde determinant with an exponentially
decaying factor in the $s$-coordinates, as in the Laughlin wave function. We still
need to check that the power $N$ is related to the denominators of the fractions
in the Hall conductance, as is the case for the Laughlin wave functions.
This is satisfied in our setting, because of the relation \eqref{betajNgdim} in
the anyon representation. In fact, notice that, for the data of an orbifold vector bundle on $\Sigma$ 
the quantization of the Hall conductance is given, through the higher twisted index theorem,
by integer multiples of Stake orbifold Euler characteristics $\chi^{orb}(\Sigma)=\chi(\Sigma_{g'})/\#G$,
hence the denominator is the order of $G$, which is also the least common multiple
of the orders $\nu_j$ of the stablizers of the cone points (Lemma 7.11 of \cite{Farb}).
The relation \eqref{betajNgdim} then relates $N$ to $\# G$.

\smallskip
\subsection{Vandermonde determinants, Selberg integrals, and Euler characteristics}

The description of the Vandermonde determinant as Jacobian
of the change of coordinates on symmetric products also leads to the
well known probability distributions in random matrix theory
\begin{equation}\label{intRMT}
\int_{\cV_\F(n)} \exp(-\| y \|^2/2) dy = \kappa \int_{\R^n} V(x_1,\ldots,x_n)^\alpha\, \exp(-\sum_i x_i^2/2) dx,
\end{equation}
where $\cV_\F(n)$ is the space of $n\times n$-hermitian matrices, with
$\F$ either the real numbers, the complex numbers or the
quaternions, respectively with $\alpha=\dim_\R \F\in \{1,2,4\}$. The numerical
factor $\kappa$ can be computed explicitly using Selberg integrals, in terms of 
Gamma functions, see \cite{FaKo} p.121.

\smallskip

Selberg integrals, and expectation values with respect to the random
matrix probability distribution
$$ V(x_1,\ldots,x_n)^\alpha\, \exp(-\sum_i x_i^2/2) dx ,$$
play a crucial role in the Harer--Zagier computation of the orbifold
Euler characteristic of the moduli spaces $\cM_{g,n}$ of algebraic 
curves of genus $g$ with $n$ marked points, \cite{HaZa}. More recently,
a {\em parameterized Euler characteristic} of $\cM_{g,n}$ was introduced
in \cite{GouHaJa}. This depends on a continuous parameter $\gamma$
and interpolates between the case of complex and real curves, respectively
corresponding to $\gamma=1$ and $\gamma=1/2$. The parameterized 
Euler characteristic is expressed in \cite{GouHaJa} in terms of Jack 
symmetric functions and of expectation values 
\begin{equation}\label{expRMTgamma}
\langle f(\lambda) \rangle = \frac{\int_{\R^n} f(\lambda)\, |V(\lambda)|^{2\gamma} \,\, e^{-\frac{\gamma}{2} p_2(\lambda)}  d\lambda}{\int_{\R^n} |V(\lambda)|^{2\gamma} \,\, e^{-\frac{\gamma}{2} p_2(\lambda)}  d\lambda},
\end{equation}
with $p_2(\lambda)=\sum_{i=1}^n \lambda_i^2$. More precisely, the parameterized 
Euler characteristic is explicitly computed in \cite{GouHaJa} in terms
of Selberg integrals of the form
$$ \int_{\R^n} |V(\lambda)|^{2\gamma} \,\, \prod_{j=1}^n (1+i\frac{\lambda_j}{a})^{-\alpha}  \,\,
(1-i\frac{\lambda_j}{b})^{-\beta} \,\, d\lambda, $$
which again can be computed explicitly in terms of Gamma functions.

It would be interesting to see if Laughlin type wave functions would arise in 
analogous computations of orbifold Euler characteristic of moduli spaces of
good $2$-dimensional orbifolds. Notice that a Teichm\"uller theory for orbifold
was developed in \cite{Thur}, see also \cite{Choi}.

\subsection*{Acknowledgments}
The first author is supported by NSF grants
DMS-1007207, DMS-1201512, PHY-1205440. 
The second author contributed to this project as part of his summer 
undergraduate research.

\end{document}